\documentclass[%
 reprint,
 amsmath,amssymb,
 aps,
showkeys,
]{revtex4-2}

\usepackage{graphicx} 
\usepackage{dcolumn} 

\usepackage{comment}
\usepackage[dvipsnames]{xcolor}

\usepackage{bm}
\newcommand{\bb}{\bm{b}}
\newcommand{\bber} {\textrm{BER}^{b}}
\newcommand{\bberi}{\textrm{BER}^{b}_{\textrm{in}}}
\newcommand{\bbero}{\textrm{BER}^{b}_{\textrm{out}}}

\usepackage{siunitx}
\DeclareSIUnit\Mbps{Mbps}
\DeclareSIUnit\GSa{GSa}
\DeclareSIUnit\dBm{dBm}
\newcommand{\ber}[1]{$10^{#1}$}

\usepackage{caption}
\usepackage{booktabs} 

\usepackage{subcaption}
\usepackage{hyperref}
\usepackage[capitalise]{cleveref}

\begin{document}


\title{\texorpdfstring{
A Microring as a Reservoir Computing Node:\\Memory/Nonlinear Tasks and Effect of Input Non-ideality
}{
A Microring as a Reservoir Computing Node: Memory/Nonlinear Tasks and Effect of Input Non-ideality
}}

\author{Davide Bazzanella}%
\email{davide.bazzanella@unitn.it}
\author{Stefano Biasi}
\author{Mattia Mancinelli}
\author{Lorenzo Pavesi}
\affiliation{ 
    Nanoscience laboratory, Department of Physics, University of Trento,
    Via Sommarive 14, 38123, Trento, Italy
}


\date{\today}

\begin{abstract}
The nonlinear response of an optical microresonator is used in a time multiplexed reservoir computing neural network. Within a virtual node approach combined with an offline training through ridge regression, we solved linear and nonlinear logic operations.  
We analyzed the nonlinearity of the microresonator as a memory between bits and/or
as a neural activation function. This is made possible by controlling both the distance between bits subject to the
logical operation and the number of bits supplied to the ridge regression. 
We show that the optical microresonator exhibits up to two bits of memory in linear tasks and
that it allows solving nonlinear tasks providing both memory and nonlinearity. 
Finally, we demonstrate that the virtual node approach always requires a comparison
of the reservoir’s performance with the results obtained by applying the same
training process on the input signal. 

\end{abstract}

\maketitle

\section{Introduction}
\label{sec_introduction}
Nowadays, artificial neural network (ANNs) are able to carry out tasks of
remarkable complexity and abstraction.
These statistical models are usually created by analysing extensive datasets
or by repeating a training procedure a substantial number of times.
Thanks to the recent exponential increase in computational power of general
purpose accelerators, derived from graphic processing unit (GPU) architectures,
it has been possible to create and emulate ANNs of increasing size, in a shorter
time and more efficiently \cite{thompson2020computational, chen2014big}.
However, a fundamental difference between ANNs and common computers still remains:
in the former, computation occurs in parallel at each node, while in the latter
information is elaborated in CPUs and GPUs only.
Due to this discrepancy, computers are not the most energy efficient means
for creating and running ANNs.

In order to overcome these limitations, hardware implementations of ANNs with
different combinations of architectures and physical substrates have been
attempted
\cite{furber2016large, nakane2018reservoir, sunada2021photonic, du2017reservoir, merolla2014million, guo2019polarization,wu2021programmable,jaeger2001echo,maass2004computational}. 
Reservoir computing (RC) with photonics is an interesting approach to the matter,
as it combines the bandwidth, parallelism, and low energy consumption proper
of photonics and the weak requirements of RC based ANNs \cite{Survey}.
In these ANN, a recurrent untrained network of randomly interconnected nonlinear neurons (\textit{the reservoir}) is interfaced through trainable connections with a single layer of output neurons (\textit{the output perceptron}). In photonics, the \textit{reservoir} that increases the dimensionality of the RC
input data can be implemented either by a big number of simpler nodes \cite{IEEE_Bienstman} or
by a single much more complex and highly nonlinear element \cite{appeltant2011information}.
This is especially important for integrated photonics, whose fundamental blocks
have limited cascadability.

An approach to photonic RCs aims at using a single photonic device
to create the \textit{reservoir}, whose single output is sampled at
different times, creating a set of virtual nodes. This method is called
time multiplexing, and require memory storage of the virtual nodes. 
Often, the output layer is carried out offline, after the optical
readout has been converted to digital signals
\cite{sunada2021photonic, appeltant2011information, donati2021microring}.
The final network output is computed as a weighted sum of the status of the virtual nodes.
Therefore, this RC implementation can be easily trained by using a simple ridge regression.
This procedure allows the study of promising optical reservoir, without
the additional burden of integrating the output regression layer.

Within this approach particular care should be placed on the network testing phase when binary inputs are used. In fact, the optical binary input signal differs from the ideal binary input
because of the non-ideal response of the optical bit generation stage, usually
composed by a  continuous wave (CW) laser and one or more electro-optic (E/O) modulators. For this reason, as we will show in this article, it is fundamental to compare the performance of the RC system to that of the isolated readout layer
applied directly to the optical input.

In this work, we studied the implementation of a silicon microresonator as a reservoir. Referring to our recent work \cite{Borghi_2021}, a binary time sequence of bits is injected into the system at different bitrates, average powers, and detuning with respect to the resonant frequency.
As a result, the nonlinearity of the microresonator, which follows the free
carrier and temperature dynamics \cite{borghi2020OntheModeling}, encode the information in the output response. The free carrier recombination and the temperature cooling provide the intrinsic fading memory to the network. Note that here, at difference with \cite{Borghi_2021}, we use a single pulsed input optical signal.
Training the system by means of an off-line ridge regression, allows treating different
binary tasks, isolating the role of the nonlinearity of the microresonator on
the fading memory between bits and on the nonlinearity (activation function)
imprinted on the output response.

The paper is organized as follows.
To begin with, we present in \cref{sec:principle} the basic principle of the network,
explaining the implementation of a microresonator/bus waveguide system as a reservoir.
We describe the encoding of information into the input signal and the training process.
We then present the experimental realization in \cref{sec:experimental}, where we
discuss the samples and the experimental setup, showing how the data were acquired.
In \cref{sec:result} we show the experimental results obtained in the test process,
discussing the role of the nonlinearity induced by the microresonator.
Finally, we summarise our main results in \cref{sec:conclusion}.

\section{Theory and basic principle}
\label{sec:principle}

A single microresonator in the add-drop configuration is used as a reservoir,
implementing virtual nodes through the time multiplexing technique \cite{Memory_OSA}.
The number of virtual nodes $N_v$ is determined by the bit duration $T_b$ and the virtual node temporal separation
$(\delta t)$, i.e. $N_v = T_b / \delta t$.
Note that in this case, $T_b$ is determined by the used bit rate and is not connected to any delay-loop as in the
classical architectures \cite{Time_Fischer, NatCom2014}.

The input signal is an optical binary sequence $\bm{I}_N$ of length $N$, encoding the
logical binaries 0 and 1 with the lowest and maximum optical intensities of a single frequency pump laser,
respectively.
The input optical signal is then a sequence of bits $\bb_j^i$, each of which is composed
by a $N_v$ number of samples $i_j$, or virtual nodes:
\begin{align}
    \label{eq:input-sequence-bits}
    \bm{I}_N &= \left( \bb_1^i, \,\dots, \bb_j^i, \,\dots, \bb_N^i \right) \\
    \label{eq:input-bit-nodes}
    \bb_j^i  &= \left( i_1, \,\dots, i_j, \,\dots, i_{N_v} \right) .
\end{align}

The possibility of varying both the frequency and power of the pump laser allows to study the effects of the microring resonator nonlinear response \cite{borghi2020OntheModeling}.
This nonlinearity, applied to the input signal, is due to either or both the dynamics of the free carrier population density and the temperature within the ring's waveguides.
The pump laser generates free carriers through Two Photon Absorption (TPA) which, in turn via free carrier dispersion, generates a blue shift of all the resonant frequencies of the microresonator.
On the other hand, the temperature of the microresonator increases due to free carrier relaxation and light absorption in the waveguide material. 
This induces a red shift of all the resonant frequencies due to the thermo-optic effect.
The two effects are characterized by different relaxation times and power dependences.
Then, depending on the frequency detuning (difference between the pump laser frequency and the microring resonant frequency), the input power and the bit rate, one of these phenomena can overcome the other or both can occur leading to an unstable scenario characterized by a self-pulsing regime \cite{johnson2006self, self_Baker}. As a result, such nonlinearities define the connection between the virtual nodes \cite{Borghi_2021}, providing both the memory capability and the nonlinearity to the reservoir.

The output bit $\bb_j^{o}$ is measured at the output port of the system, i.e. the microresonator transmitted light is detected and sampled each $\delta t$ yielding the virtual nodes status $o_j$. These output bits are arranged in the hidden nodes matrix:
\begin{align}
    \label{eq:output-sequence-bits}
    \bm{X}  &= \left( \bb_1^{oT}, \,\dots, \bb_j^{oT}, \,\dots, \bb_N^{oT} \right) \\
    \label{eq:output-bit-nodes}
    \bb_j^o &= \left( o_1, \,\dots, o_j, \,\dots, o_{N_v} \right),
\end{align}
where each column $\bb_j^o$ contains $N_v$ virtual nodes. Then, offline, the RC output $\bm{Y}$ is obtained by a simple matrix multiplication of the hidden nodes matrix with a weight matrix $\bm{W}$, i.e. $\bm{Y}= \bm{X} \bm{W}$

The reservoir training is obtained by determining the weight matrix $\tilde{\bm{W}}$, which allows predicting the target $\bm{Y_T}$. This problem is solved by regularized least squares (ridge regression), exploiting the \texttt{fitrlinear} algorithm of Matlab.
Here, the regularization parameter $\lambda$ is defined by a 5-fold cross validation, so that the result of the ridge regression is the matrix $\tilde{\bm{W}}$ which minimizes the regularized least square error.

\begin{table}[t!]
    \centering
    \begin{tabular}{c|cc}
        AND & 0 & 1 \\
        \hline
          0 & 0 & 0 \\
          1 & 0 & 1
    \end{tabular}
    \qquad
    \begin{tabular}{c|cc}
         OR & 0 & 1 \\
        \hline
          0 & 0 & 1 \\
          1 & 1 & 1
    \end{tabular}
    \qquad
    \begin{tabular}{c|cc}
        XOR & 0 & 1 \\
        \hline
          0 & 0 & 1 \\
          1 & 1 & 0
    \end{tabular}
    \caption{ \label{tab:truth-tables} Truth table for AND, OR, and XOR operations.}
\end{table}

\begin{figure}[!ht]
    \centering
    \includegraphics[scale = 1]{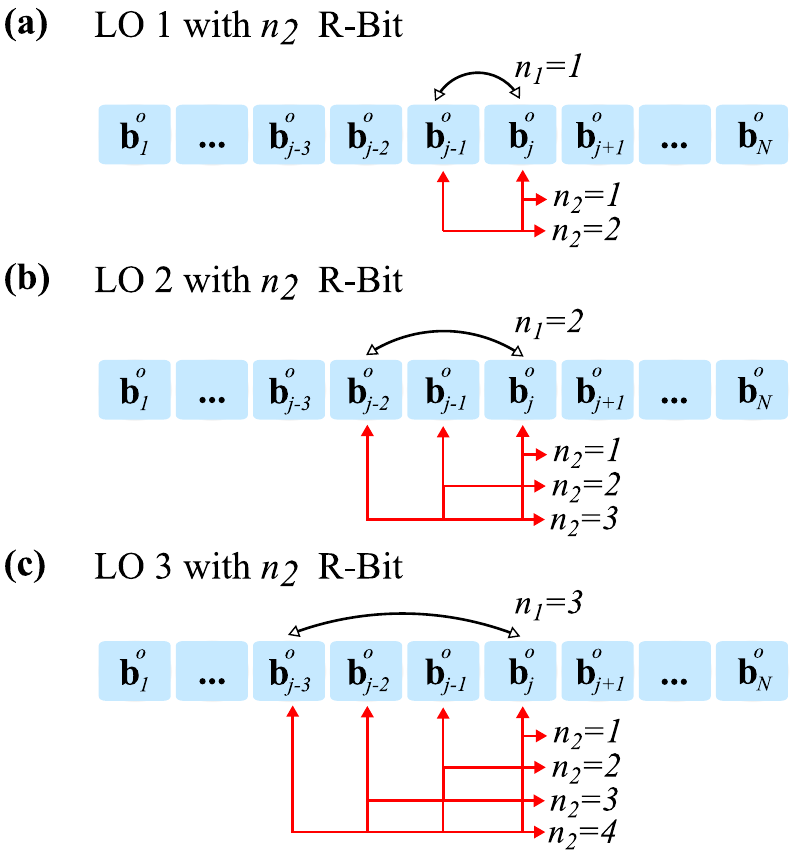}
    {\phantomsubcaption \label{fig:bit_register1}}
    {\phantomsubcaption \label{fig:bit_register2}}
    {\phantomsubcaption \label{fig:bit_register3}}
    \caption{ \label{fig:bit_register}
        Sketch representing the three cases on which we tested the logical operations.
        $n_1$ indicates the distance between the bits on which the logical operation (LO) is performed and $n_2$ is the number of bits provided to the ridge regression in the training procedure. Note that the flow of bits is such that the past bits ($\bb_{j-n_1}^i$) are processed by the microresonator before the present bit ($\bb_{j}^i$), i.e. the bit flow is inverted with respect to the time flow which is indicated in the figure.
    }
\end{figure}

The binary tasks studied are the logical operations AND, OR, and XOR, whose truth tables are reported in \cref{tab:truth-tables} and which are carried out on the present bit with a bit in the past. To explicit the bits on which the operation is carried out, we use the following notation: ``LO $n_1$ with $n_2$ R-bit'', where LO is the logical operation, $n_1$ is the distance between the bits on which the LO is performed (the present and the past bits), and $n_2$ is the number of bits, starting from the present one, provided to the ridge regression (the R-bits).
For example, ``AND 2 with 2 R-bits'' means to perform an AND logical operation between the $\bb_j^i$ and $\bb_{j-2}^i$, providing the virtual nodes of $\bb_j^o$ and $\bb_{j-1}^o$ to the ridge regression algorithm.
For each task the following cases were investigated:
\begin{itemize}
  \item logical operation between bits $\bb_j^i$ and $\bb_{j-1}^i$, providing to the ridge regression the bit $\bb_j^o$ or $\bb_{j-1}^o\, \& \, \bb_{j}^o$, see \cref{fig:bit_register1};
  \item logical operation between bits $\bb_j^i$ and $\bb_{j-2}^i$, providing to the ridge regression the bit $\bb_j^o$ or $\bb_{j-1}^o\, \& \, \bb_{j}^o$ or $\bb_{j-2}^o\, \& \, \bb_{j-1}^o\, \& \, \bb_{j}^o$, as shown in \cref{fig:bit_register2};
  \item logical operation between bits $\bb_j^i$ and $\bb_{j-3}^i$, providing to the ridge regression the bit $\bb_j^o$ or $\bb_{j-1}^o\, \& \, \bb_{j}^o$ or $\bb_{j-2}^o\, \& \, \bb_{j-1}^o\, \& \, \bb_{j}^o$ or $\bb_{j-3}^o\, \& \, \bb_{j-2}^o\, \& \, \bb_{j-1}^o\, \& \, \bb_{j}^o$, see \cref{fig:bit_register3}.
\end{itemize}   
Varying the distance between the bits subject to the logical operation allows testing the memory imprinted on the output signal by the microresonator.
We can supply to the ridge regression the current bit only, so that the system has to provide the memory to the regression, or we can make all the bits in the operation directly available to the regression.
In the last scenario, the attention can be focused on the nonlinearity of the microresonator imprinted on the output signal.

Despite the apparent simplicity of the logical operation AND, OR and XOR, these tasks unveil the main characteristics of the reservoir: the nonlinear transformation of the input and the presence of fading memory \cite{XOR_coupled-resonator,XOR_Han,Xor_tworingNew, Borghi_2021}.
AND and OR, are linear tasks, and therefore, do not require a nonlinear response for their solution.
However, in order to solve these linear tasks it is necessary for the system to provide memory between the bits.
This allows isolating the role of optical nonlinearity on the memory between the distinct bits.
Differently, the solution of the logical XOR requires both: nonlinearity of the response and memory between the different bits considered in the task.
As a result, we can distinguish between the two contributions by providing or not providing memory to the ridge regression during the training process.

\section{Samples and experimental realization}
\label{sec:experimental}

\begin{figure*}[t]
    \centering
    \includegraphics[scale = 1]{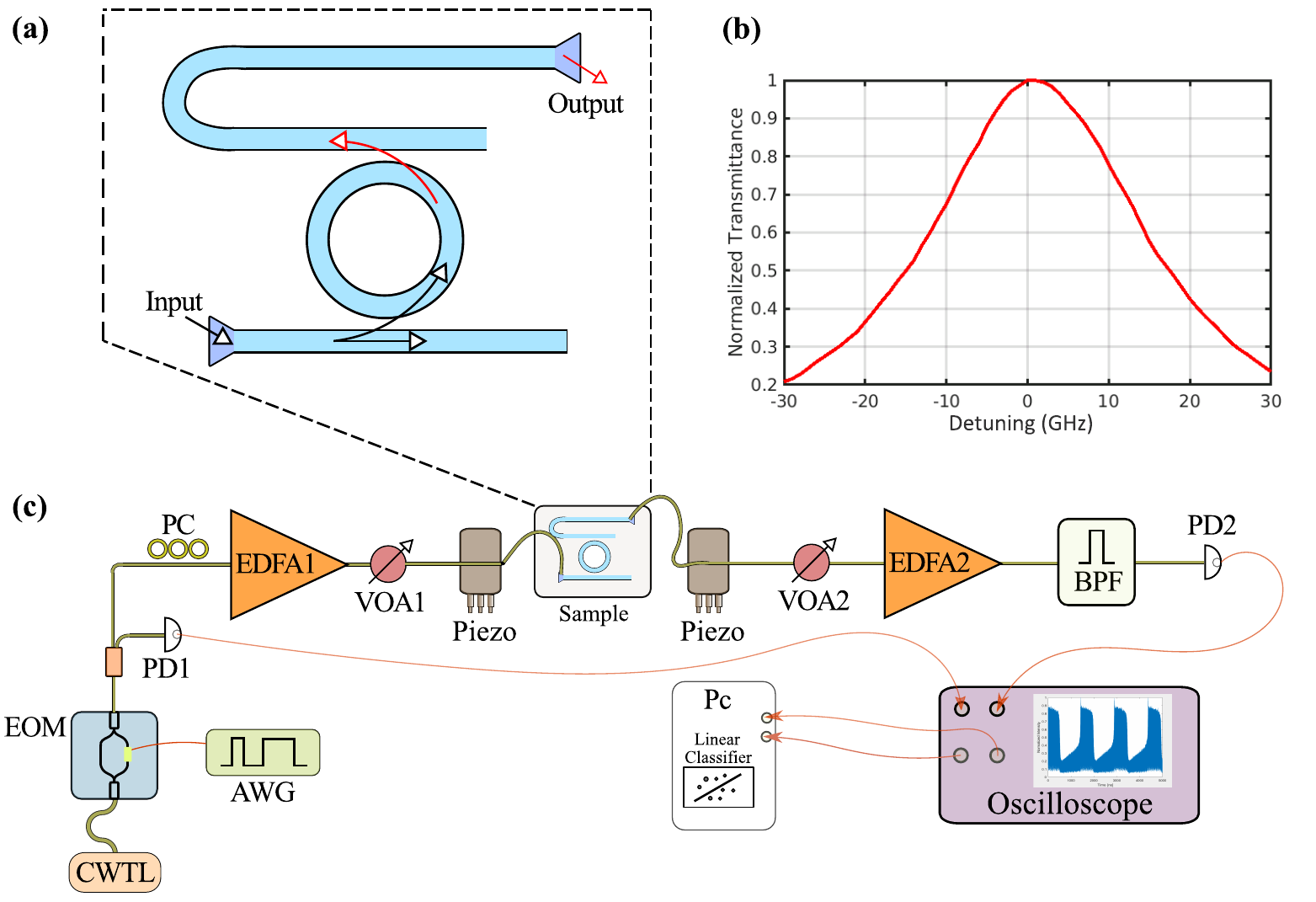}
    {\phantomsubcaption \label{fig:uring}}
    {\phantomsubcaption \label{fig:spectrum}}
    {\phantomsubcaption \label{fig:setup}}
    \caption{ \label{fig:experiment}
        (\subref{fig:uring}) Sketch of the microring resonator in the add-drop configuration.
        (\subref{fig:spectrum}) Normalized transmission spectrum around the resonance frequency. The detuning is the frequency difference between the laser frequency and the microring resonat frequency of  \qty{193.5}{\THz}.
        (\subref{fig:setup}) Diagram of the experimental setup.
            CWTL: Continuous Wave Tunable Laser,
            AWG: Arbitrary Waveform Generator,
            EOM: Electro-Optic Modulator,
            PD: Photodetector,
            PC: polarization control,
            VOA: Variable Optical Attenuator,
            EDFA: Erbium Doped Optical Amplifier,
            BPF: Band Pass Filter,
            Pc: Personal computer.
    }
\end{figure*}

The device under test (DUT) was a microring resonator with a radius of \qty{7}{\um} in the add-drop configuration, which has been fabricated at the CEA-Leti facility on a SOI (silicon-on-insulator) wafer. The top-view of the sample is shown in the sketch of \cref{fig:uring}.
Briefly, the DUT is composed by single mode channel silicon waveguides with a width and height of \qty{450}{\nm} and \qty{220}{\nm}, respectively.
Two bus waveguides are point-coupled to the microring with a gap of \qty{200}{\nm}. The normalized transmittance spectrum of a resonance measured at the drop port is shown in \cref{fig:spectrum}.
From this response it is possible to estimate a quality factor of about \num{6e3} at a frequency of \qty{193.5}{\THz}.
As a result, the photon lifetime in the cavity is equal to $\qty{4.93}{\ps}$ \cite{Time_Biasi}.
Differently, the free carrier lifetime $\tau_{FC}$ is about a few nanosecond, as reported in literature for similar structures \cite{1.9ns_Pernice, xu2007all, almeida2004all, luo2012power}, while the thermal relaxation time is about $\qty{100}{\ns}$ \cite{Thermal_Van, Thermal_Van1}.
It is worth noting that this $\tau_{FC}$ is about one order of magnitude shorter than the value estimated in our works \cite{borghi2020OntheModeling, Borghi_2021}, where $\tau_{FC} = \qty{45}{\ns}$ on IMEC fabricated microresonators.
This means that the transient phenomena occur at a temporal scale 10 times shorter in our DUT with respect to what was found in \cite{Borghi_2021}.
Indicatively, at \qty{8}{\dBm} the response of these DUTs can be considered linear at all frequencies.
Increasing the average input optical power, the nonlinear effects become more significant.
At the highest values of input power, earlier for small detuning values and later for larger detunings, self-pulsing happens.
For example, an input signal in resonance with the ring with an average power of \qty{16}{\dBm} triggers the self-pulsing effect.
The possibility of varying the frequency of the pump laser from negative to positive detunings, allows forcing a particular nonlinear effect.
In fact, negative detuning induces free carriers through TPA, while positive detuning favors the thermo-optic effect.
However, with a fixed input frequency at high input power, the microresonator exhibits a nonlinearity that is effectively an interplay between the two effects \cite{Borghi_2021}.

The experimental setup is sketched in \cref{fig:setup}.
The pump is a CW tunable laser, which operates in a range spanning from \qtyrange{191.5}{196.25}{\THz} (CWTL).
Its intensity is modulated by an electro-optic IQ modulator (EOM), controlled by a \SI[per-mode = symbol]{65}{\GSa\per\second} Arbitrary Waveform Generator (AWG), in order to create the desired binary sequence.
This signal passes through a fiber-optic splitter which allows detecting the input pump by a fast photodiode detector (PD1).
The remaining of the pump is adjusted in polarization by a polarization control (PC), is amplified by an Erbium Doped Optical Amplifier (EDFA1), and attenuated again to obtain the required power by an electronic Variable Optical Attenuator (VOA1).
The resulting signal is fed to the sample through the coupling between a single mode fiber and the input grating coupler.
Similarly, another single mode fiber is coupled to the output grating coupler to read the output signal.
A correct alignment is ensured by a three axis linear piezoelectric stage on both the input and output.
The signal collected at the output is injected into a second electronic Variable Optical Attenuator (VOA2), re-amplified by a second Erbium Doped Optical Amplifier (EDFA2), and cleaned from the optical ,noise with a tunable band-pass filter (BPF).
The resulting output is detected by a second photodiode (PD2).
At the end, a 4-channel \SI[per-mode = symbol]{40}{\GSa\per\second} oscilloscope monitors and records the input and output optical waveform.
A Personal computer (Pc) elaborates the traces, and performs an offline training and test of the network by using Matlab.

The input signal injected into the microresonator consists of a Pseudo Random Binary Sequence (PRBS) of order 8 and length 255, repeating indefinitely.
Its characteristics can be changed by varying three distinct variables: bitrate, detuning (input minus resonant frequency) and average power.
Precisely, the variables span over the following values: 
\begin{itemize}
    \item bitrate: \SIlist{20;40;50;80;100;200;250;400;500;800;1000;2000;4000}{\Mbps};
    \item detuning: \SIlist{-30;-25;-20;-15;-10;-5;0;5;10;15;20;25;30}{\GHz};
    \item input Power: \SIlist{8;9;10;11;12;13;14;15;16;17;18}{\dBm}.
\end{itemize}
For each combination of bitrate, incident power, and frequency detuning, the input and the output optical signals are acquired.
The second amplification stage, consisting of the EDFA2 and of the VOA2, keeps constant the average power at PD2, avoiding marked changes in the signal-to-noise ratio (SNR), which could otherwise compromise the performance comparison between different combination of parameters.

In the time multiplexing approach, knowledge of the input is crucial to verify that the experimental apparatus alone is not able to solve the task.
Indeed, both the signal generation and detection stages can distort the ideal PRBS by imprinting spurious nonlinearities and adding unwanted memory due to their finite electronic bandwidth.
Therefore, the oscilloscope records both the input (the PD1 signal) and the output (the PD2 signal) optical waveforms, with a fixed sampling rate of \SI[per-mode = symbol]{20}{\GSa\per\second}.
As a result, the sampling rate of the oscilloscope defines the number of samples in each bit for each bitrate.
Specifically, it spans from a maximum of 1000 to a minimum of 5, for \qty{20}{\Mbps} and \qty{4000}{\Mbps}, respectively.

The acquired samples per bit are re-binned to obtain $N_{v}^{d}$ virtual nodes: they are divided into the desired number of bins and for each ones is performed the average.
For example, assuming a $N_{v}^{d}=10$, each bit would have 10 virtual nodes obtained as follows:
\begin{enumerate}
	\item 10 samples obtained by re-binning the experimental data, at input bitrates from \qtyrange{20}{1000}{\Mbps};
	\item all the 10 samples acquired, at \qty{2000}{\Mbps};
	\item the values of the five samples acquired and zero for the remaining 5 nodes, at \qty{4000}{\Mbps}.
\end{enumerate}

The minimization of the regularized square error, i.e. the solution of the system $\bm{Y_T}=\bm{X}\tilde{\bm{W}}$, is performed on both the input and the output signals by processing them as a function of the maximum desired number of virtual nodes ($N_{v}^{d}$).

\section{Experimental results}
\label{sec:result}

For each combination of the three input variables (see \cref{sec:experimental}), we performed the training for different logical operations on both the input and the output optical signals, ie. we detected the optical data and then we applied the ridge regression on the digital data. When the training is performed on the output optical signal we are using the whole microring based RC network. We studied a number of virtual nodes equal to \numlist{3;4;5;10;15;20;30}. The tasks have been implemented following the cases reported in \cref{sec:principle}. Then, the performance of the network has been assessed by estimating the bit error rate (BER).

Among the several results, we selected a few instructive cases, which reveal the effect of the nonlinearity of the microring resonator.
We report the results separating them according to the logical operation.
In each case, we show three contour maps as a function of the input bitrate and the frequency detuning. The first map shows the best value of the BER ($\bbero$) obtained by the RC network with the input power which ensures the best performance. The second map shows the lowest value of input power at which $\bbero$ is achieved.
The third one shows the ratio $RB$ between the $\bber$ estimated when the ridge regression is applied on the input optical data ($\bberi$) or on the output optical data ($\bbero$): $RB = \bberi/\bbero$.
The color code for the ratio $RB$ is given on the map with two different color-ranges: one from blue to yellow, the other from black to white.
The first highlights where the microring introduces an improvement of the performance. Specifically, the yellow indicates a better result of the RC network, and therefore, defines the regions of performance improvement given by the microring nonlinearity. The other color range is a gray scale that shows where the results of the RC are worse than barely performing the ridge regression on the input optical data.
In particular, the black color indicates that $RB$ is equal to one, i.e. the task is solved with the same BER either when the input optical data are processed or when the output optical data are processed, while the gray and white colors indicate regions in which $\bbero$ is worse than $\bberi$. This means that the nonlinear response of the microring does not introduce any advantage but it is only detrimental.
For the sake of clarity, all the maps are represented with a logarithmic ($\log_{10}$) scale.

Moreover, in the map of the $\bbero$ we define with red dots the points where the results reach the statistical limit. The same points are replicated in the third map, using empty red dots and crosses when the statistical limit is reached by processing the input and output optical signals, respectively.

\subsection{Linear logical operations: AND and OR}
\label{ssec:AND-OR}

Logical AND and OR tasks are linearly separable (see \cref{tab:truth-tables} for the truth tables), hence the nonlinearity of the microresonator provides only the memory of the past bit for the operation.
As a result, varying the distance between the bits subject to the logical operation allows testing the memory capability of the network.

\subsubsection{AND 1 with 2 R-bits and \texorpdfstring{$N_{v}^{d}=5$}{Nvd = 5}}
\label{sssec:AND-1-2}

\begin{figure}[t]
    \centering
    \includegraphics[scale = 1]{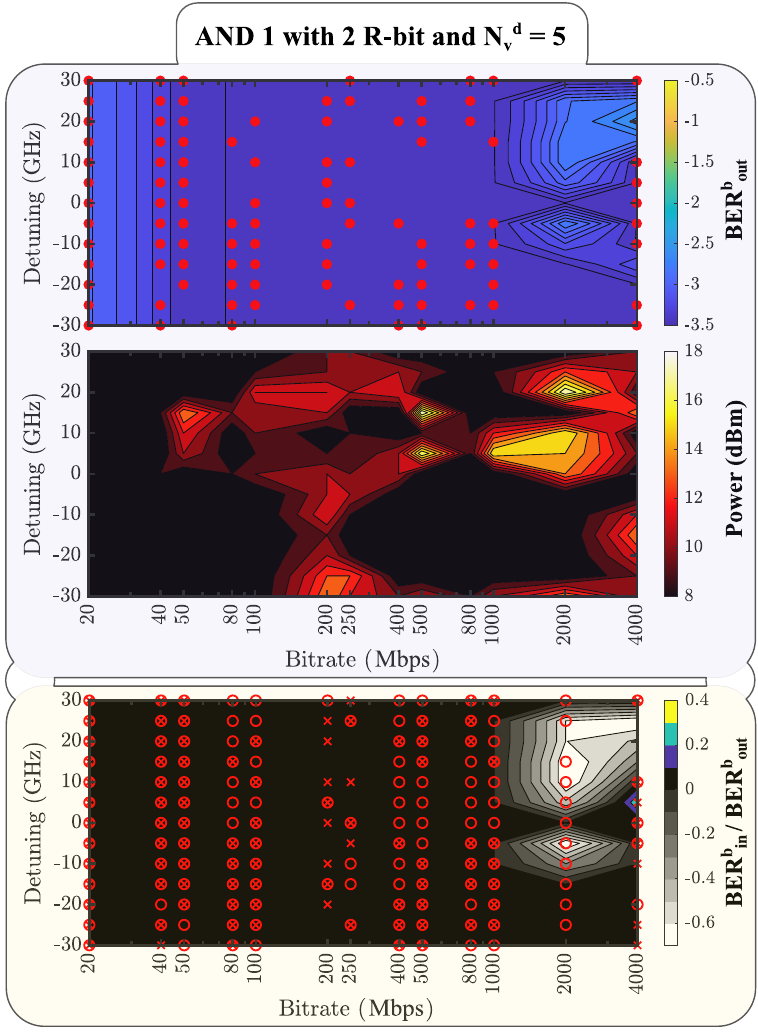}
    \caption{ \label{fig:AND1_2Rb_5Nv}
        Maps as a function of the frequency detuning and input bitrate for AND 1 with 2 R-bit and $N_v^d = 5$.
        (\textbf{top}) BER estimation from the RC network at the power which ensures the best network performances;
        (\textbf{middle}) the power at which the $\bbero$ values in the first panel are achieved;
        (\textbf{bottom}) the ratio between $\bberi$ and $\bbero$. All the values are given in a logarithmic scale.
    }
\end{figure}

The ``AND 1 with 2 R-bits'' is a linear task where we provide both bits - the current bit and the past one - to the ridge regression.
In this case, we expect to be able to solve the task both by the RC and by processing only the input optical signal. The results confirm these expectations and are shown in \cref{fig:AND1_2Rb_5Nv}.

The system is in fact able to solve the task without errors in almost all configurations, as we can see from the top panel.
However, we can observe a slight deterioration of the the BER in some isolated regions, probably due to lower SNR (signal to noise ratio).
The absolute minimum $\bber$ value is equal to \ber{-3.4} and the statistical limit is achieved at all input bitrates apart from \qty{2000}{\Mbps}.
Looking at the power map (see middle panel), there is no clear dependence of the $\bber$ on the frequency detuning.
Furthermore, the best BER results are found at low values, whereas, the lowest performance is characterized by high powers.
As expected, the bottom map shows that the performance of the RC is equal to that achieved by processing only the input optical signal almost everywhere.
For bitrates around \qtylist{2000;4000}{\Mbps} $\bbero$ is even worse than $\bberi$.
In these cases, the microresonator distort the information carried by the input and it is, therefore, detrimental to the task resolution.

\subsubsection{AND 1 with 1 R-bit and \texorpdfstring{$N_{v}^{d}=5$}{Nvd = 5}}
\label{sssec:AND-1-1}

\begin{figure}[t]
    \centering
    \includegraphics[scale = 1]{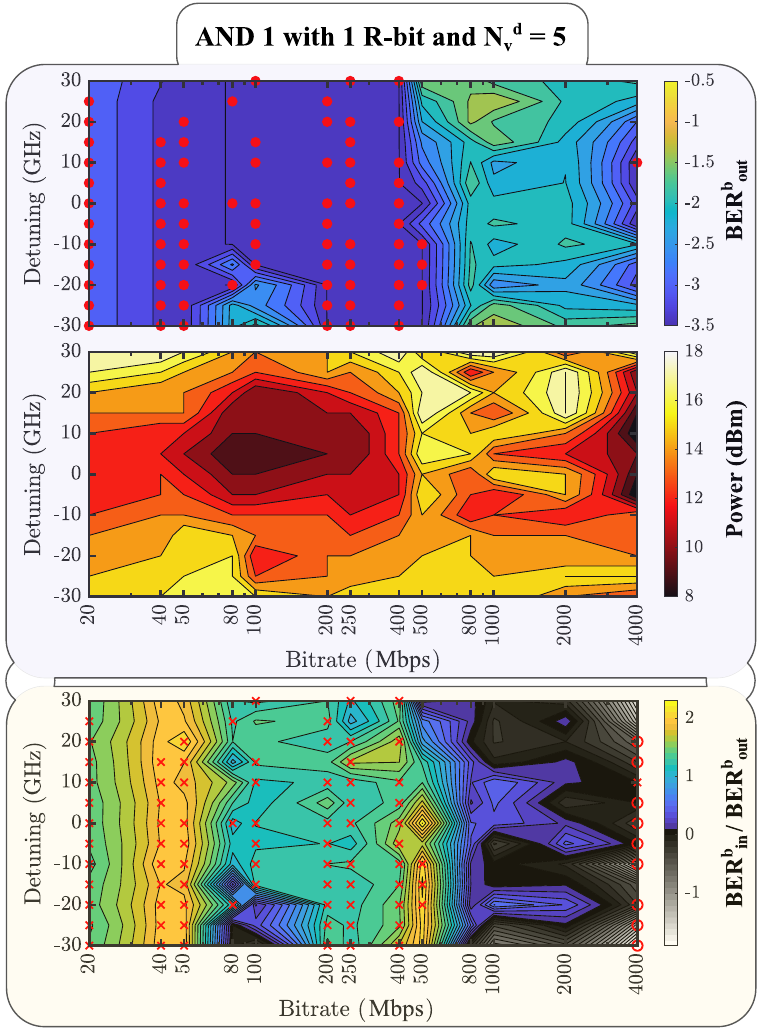}
    \caption{ \label{fig:AND1_1Rb_5Nv}
        Same maps of \cref{fig:AND1_2Rb_5Nv} for AND 1 with 1 R-bit and $\protect N_v^d = 5$.
    }
\end{figure}
\begin{figure}[t]
    \centering
    \includegraphics[scale = 1]{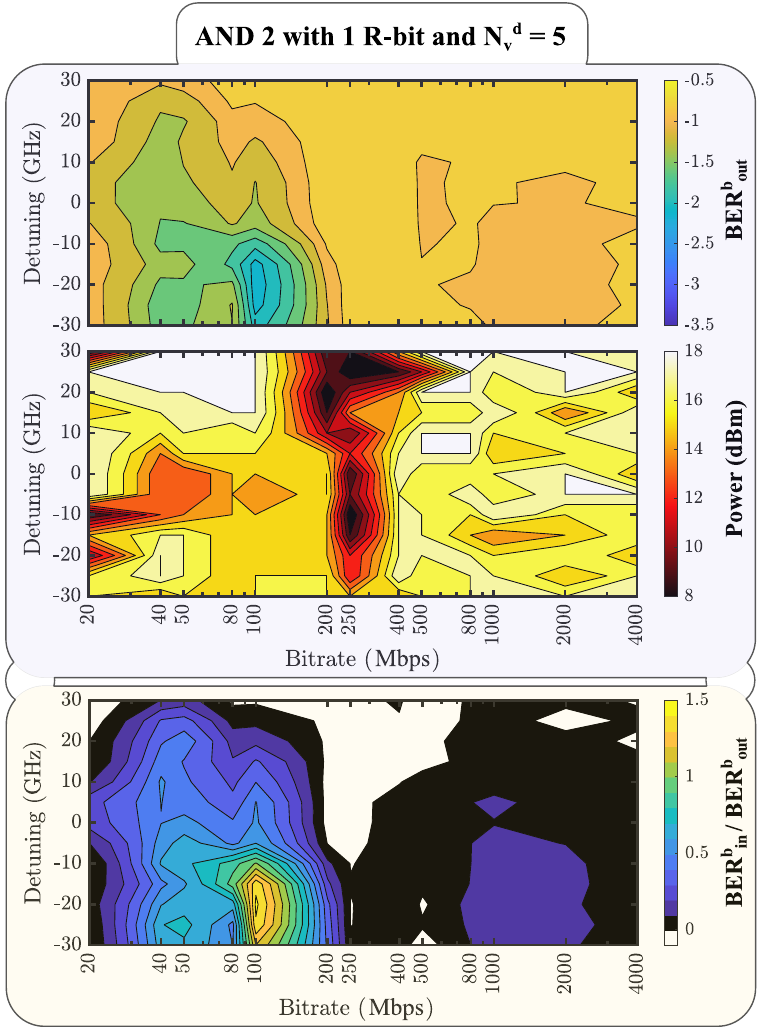}
    \caption{ \label{fig:AND2_1Rb_5Nv}
        Same maps of \cref{fig:AND1_2Rb_5Nv} for AND 2 with 1 R-bit and $\protect N_v^d = 5$.
    }
\end{figure}

Similarly to the previous example, ``AND 1 with 1 R-bits'' is a linear task.
However, differently than before, we provide now only the current bit to the ridge regression.
The nonlinearity of the microring resonator, then, must provide memory of the past bit to the regression.
Indeed, the past bit has been injected into the microring resonator before the current bit and should be stored in the microring resonator, as shift in temperature or free carrier population, to influence the current bit transmission. 
The results are shown in \cref{fig:AND1_1Rb_5Nv}.

The best BER estimated (top panel) shows low values for a vast portion of the bitrate-detuning parameter space, with an absolute minimum of \ber{-3.4}.
Moreover, the statistical limit (red dots) is reached at several frequency detuning values for bitrates ranging from \qty{20}{\Mbps} to about \qty{500}{\Mbps}.
From this map alone we cannot infer a clear dependence of the $\bbero$ as a function of the frequency detuning.
On the contrary, the power at which $\bbero$ is obtained (see middle panel) shows that at zero detuning the system solves the task even at low power, while away from the resonance it requires higher incident powers.
Lastly, from the bottom panel, showing the ratio $RB$, we can observe wide regions of marked improvement in the performance of the RC network with respect to processing only the input optical signal.
This improvement extends roughly from \qtyrange{20}{500}{\Mbps} and reaches about two orders of magnitude in the yellow region between \qtylist{40;50}{\Mbps}.
On the other hand, for higher bitrates (approximately from \qtyrange{1000}{4000}{\Mbps}) the results provided by the RC network equals those obtained by processing only the input optical signal (dark gray and black regions).
The extreme case occurs at \qty{4000}{\Mbps}, where the ridge regression achieves the statistical error limit on processing the input optical signal, only.
This confirms what we have seen in the previous case, where at high input bitrates the nonlinearity distort the signal, and therefore, it does not play a memory role.
Furthermore, this verifies the presence of unwanted memory in the signal, due to either the generation or the detection stages, even capable, in this extreme case, of solving the task.


\begin{figure}[b]
    \centering
    \includegraphics[scale = 1]{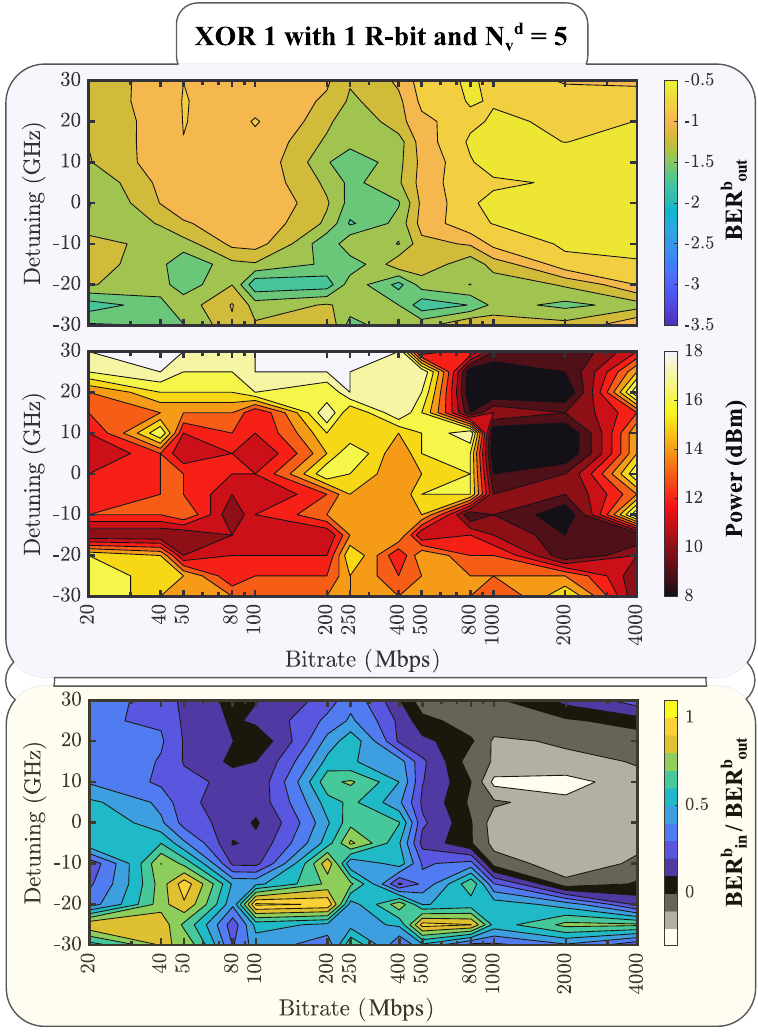}
    \caption{ \label{fig:XOR1_1Rb_5Nv}
        Same maps of \cref{fig:AND1_2Rb_5Nv} for XOR 1 with 1 R-bit and $\protect N_v^d = 5$.
    }
\end{figure}
\begin{figure}[b]
    \centering
    \includegraphics[scale = 1]{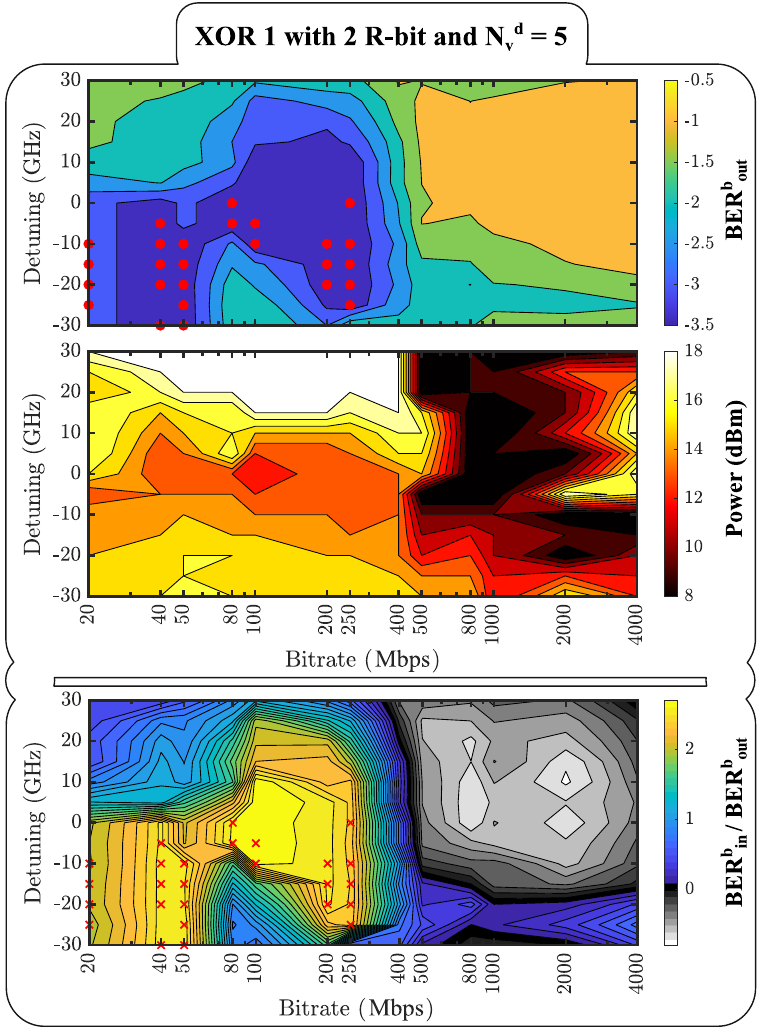}
    \caption{ \label{fig:XOR1_2Rb_5Nv}
        Same maps of \cref{fig:AND1_2Rb_5Nv} for XOR 1 with 2 R-bit and $\protect N_v^d = 5$.
    }
\end{figure}

\subsubsection{AND 2/3 with 1 R-bit and \texorpdfstring{$N_{v}^{d}=5$}{Nvd = 5}}
\label{sssec:AND-2/3-1}

By increasing the distance between the bits on which the logical AND operation is performed, more memory capacity of the reservoir is required.
\Cref{fig:AND2_1Rb_5Nv} shows the results for AND 2 with 1 R-bit and $N_{v}^{d}=5$.

In this case, the map of the $\bbero$ (top panel) exhibits a marked region where the RC network solves the task. In particular, there is a minimum BER of about \num{1e-2} at a bitrate of \qty{100}{\Mbps} and a frequency detuning equal to \qty{-20}{\GHz}. The map in the middle panel of \cref{fig:AND2_1Rb_5Nv} shows that this value is obtained at powers which ensure a nonlinear response of the microring resonator.
It is worth noting that the $RB$ ratio map (bottom panel) also highlights this area of better performance.
Specifically, the maximum performance improvement shows a ratio of about \ber{1.5} at a bitrate of \qty{100}{\Mbps}.
It seems that the best value of the RB occurs in the presence of a nonlinear effect related to free carrier dynamics.
In fact the best performance of the output is obtained around \qty{100}{\Mbps} at negative detuning with approximately \qty{15}{\dBm}.

Moreover, there is a small region where the RC network performs slightly better than the bare processing of the input optical signal for bitrates of \qtylist{1000;2000}{\Mbps}, but the improvement is extremely small (\ber{0.1}).
It is important to notice that the area of lower RB coincides to an area of low input power.
This suggests that in these cases the system nonlinearities are detrimental.

From these results, we find that the microring resonators used as a reservoir exhibits a memory of two bits. Interestingly, increasing the number of virtual nodes, the regions where the $\bbero$ is lowest remain unchanged while the minimum $\bber$ value progressively decreases until it reaches a value of about \ber{3.4} for 30 virtual nodes. In this case, $RB$ gets to a maximum value of \ber{-2.6}.

In the AND 3 with 1 R-bit and $N_{v}^{d}=5$ task, the absolute value of $\bbero$ reaches a minimum value of approximately \ber{-1} at a bitrate of \qty{100}{\Mbps} and \qty{-20}{\GHz} detuning where $RB$ equals to \ber{0.26}.
It is clear that the nonlinear response of the microresonator is not sufficient to guarantee the memory capacity needed to solve the 3-bit delay task.

\subsection{Nonlinear logical operation: XOR}

The XOR logical operation is not linearly separable and, therefore, it cannot be successfully solved using only the ridge regression.
Furthermore, the solution of the n-bit delay XOR requires also a memory capability corresponding to a n-bit delay between the input bits.

\subsubsection{XOR 1 with 1 R-bit and \texorpdfstring{$N_{v}^{d}=5$}{Nvd = 5}}

The XOR 1 operation requires both nonlinearity and memory, however, because it is carried out on two contiguous bits, the memory may come from the inter-symbolic interference already present in the input optical signal. The values of $RB$ should clarify the role of the microring in the RC network. The results are shown in \cref{fig:XOR1_1Rb_5Nv}.

The map of the best $BER$ (top panel) shows an absolute minimum value of about \ber{-1.7} 
and exhibits a region of low values extending for negative detuning frequencies across all the bitrates. Negative detuning is the region where the free carrier nonlinearities are better excited in the microring \cite{johnson2006self,Borghi_2021}.
Where thermal nonlinearities are excited predominantly and thermal bistability or self-pulsing occur, i.e. positive detuning or high input power, yields $BER$ degradation.
The power map (middle panel) exhibits relatively smooth surface, with lower values of power close to the resonance and higher values far from it.
Interestingly, the RC network solves the task with low $BER$ when the input bit rate is inversely proportional to the free carrier lifetime ($\sim$ \qty{250}{\Mbps}), even though for large input power for positive detunings.
The best $RB$ are obtained at several bitrates for negative frequency detunings.
On the contrary, for frequency detuning larger than \qtyrange{-20}{30}{\GHz} and input bitrates higher than \qty{800}{\Mbps}, we can observe a region where $\bbero$ is larger than $\bberi$, i.e. the microring resonator nonlinearites do not improve the results of the ridge regression applied on the input optical signal.
Analyzing the RB, it looks like that, by exploiting the microresonator as a memory and nonlinear activation function, there is a clear improvement in performance for negative detunings compared to the case where we just use the inter-symbolic interference and the square module of the response.
The latter scenario corresponds to analyzing only the input optical signal.

\subsubsection{XOR 1 with 2 R-bit and \texorpdfstring{$N_{v}^{d}=5$}{Nvd = 5}}

Considering the same XOR operation, but with both bits supplied to the ridge regression, we provide the memory needed to solve the operation and the reservoir has only to provide the nonlinearity. The results are shown in \cref{fig:XOR1_2Rb_5Nv}. 

In this case, the results are much better than in the XOR 1 with 1 R-bit case. The top panel shows a wide region, between \qtylist{20;400}{\Mbps}, in which the the $\bber$ reaches very low values, with an absolute minimum of \ber{-3.4}.
For bitrates of \qty{500}{\Mbps} and above, we can observe a region where the $\bber$ value does not reach low values (from \ber{-1} to \ber{-2}).
The middle panel is similarly divided in two regions, where the one associated with low $\bber$ values presents higher input power and the one associated with higher $\bber$ values presents lower input power.
It is clear that the system is able to exploit the microring nonlinearities in order to solve the task in the region of high input power, but not in the region of low input power.
In this situation, large input powers do not improve the performance calculated on the microring resonator transmitted signal.
Again, in the region where the RC network is effective in solving the task, the $RB$ shows a performance increase achieving a ratio of about \ber{2.7}.
Also for this task, we observe that negative detunings are better and allows reaching the statistical limit of the solution as highlighted by the red dots and crosses in the top and bottom panels, respectively. Thus, also in this case, free carrier nonlinearities are used by the RC network.



\subsubsection{XOR 2-3 with 1-2-3 R-bit and \texorpdfstring{$N_{v}^{d}=5$}{Nvd = 5}}

\begin{figure}[ht]
    \centering
    \includegraphics[scale = 1]{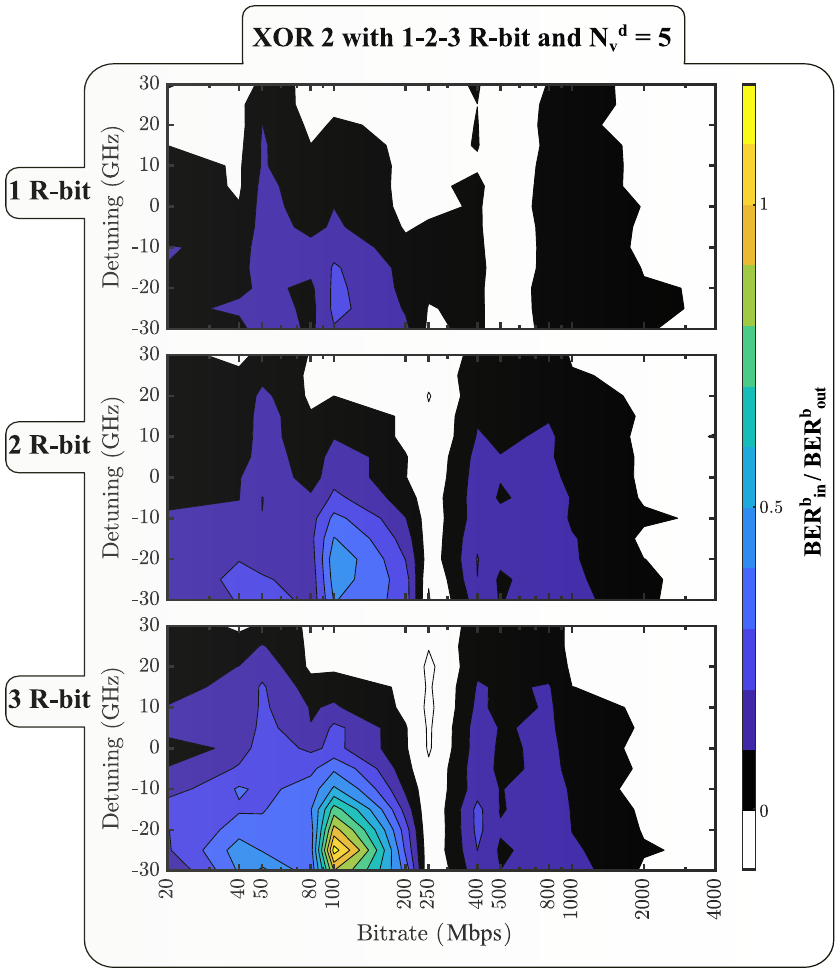}
    \caption{ \label{fig:XOR2_123Rb_5Nv}
        Map of the ratio between the input and output $\protect BER^b$ as a function of the frequency detuning and bitrate.
        Specifically, the top, middle and bottom panel shows the experimental results for $\protect N_{v}^{d} = 5$ considering the logical operation XOR 2 with 1R-bit, XOR 2 with 2R-bit and XOR 2 with 3R-bit, respectively.
    }
\end{figure}

Performing the XOR operation on bits separated by two or three bits requires even more memory than the XOR 1 operation and avoid the problem of inter-symbolic interference on the input optical signal.
\Cref{fig:XOR2_123Rb_5Nv} shows results of the XOR 2 operation by showing the map of the ratio between $\bberi$ and $\bbero$ when supplying one, two, and three bits to the ridge regression. Whenever $\bbero < \bberi$, the task is solved by the nonlinearity of the microring resonator. Note that for the 3 R-bit result, both the bits used in the XOR operation are provided to the ridge regression.

In the first case (top panel), when only the information contained in the current bit is used, the RC network is not able to improve over the bare processing of the input signal. The maximum $RB$ performance increase is \ber{0.2}. Moreover, apart from the zone contained between \qtylist{40;200}{\Mbps} and with negative frequency detuning, i.e. where the free carrier nonlinearity are effective, the map shows a slight performance decrease.

Interestingly, by providing the ridge regression also with the previous bit ($n_2 = 2$), the region showing performance increase widens and its maximum value becomes about \ber{0.4} at a bit rate of \qty{100}{\Mbps} and frequency detuning of \qty{-25}{\GHz}. Similarly as before, a second region showing good results appears for negative detuning at \qtylist{400;800}{\Mbps}. By providing all three bits to the ridge regression, the current bit and the previous two, the performance improves conspicuously reaching a maximum ratio RB of \ber{1.1} at \qty{100}{\Mbps}, \qty{-25}{\GHz}.
\begin{table}[ht]
    \centering
    \begin{tabular}{c|rrr}
      \toprule
      XOR 2 & 1 R-bit & 2 R-bit & 3 R-bit \\
      \midrule
      $\protect BER^b_{out}$ & \ber{-0.7} & \ber{-0.9} & \ber{-2.6} \\
      RB & \ber{0.2} & \ber{0.4} & \ber{1.1} \\
      \bottomrule
    \end{tabular}
    \caption{\label{tab:XOR2_123Rb_best}%
        Value of the $\protect BER^b_{out}$ and the corresponding average input optical power at 16 \unit{\dBm}, obtained for a detuning of \qty{-25}{\GHz} and a bitrate of \qty{100}{\Mbps}.
    }
\end{table}
It looks like that the network efficiently exploits only the nonlinear dynamics of the free carriers in order to solve the logical operation.
Table \ref{tab:XOR2_123Rb_best}, allows comparing the RB ratio and $\bbero$ with each other for 1,2,3 R-bit.
In particular, it reports the $\bbero$ and the RB ratio for a PRBS signal at a bitrate of \qty{100}{\Mbps}, with an average power equal to \qty{16}{\dBm}, generated by a pump laser at a frequency detuning of -25 GHz.

\subsection{Discussion}
\label{ssec:discussion}

The results of ``AND 1 with 2 R-bit'' provide us the baseline for the discussion: we verify that the ridge regression is able to correctly solve a linear task when provided with the necessary inputs.
We do this both using the input optical signal and the reservoir transmitted output optical signals. We observe that the regression reaches the statistical error limit almost everywhere.
The task ``AND 1 with 1 R-bit'' requires to carry the information of the past bit to the current one, exploiting the response of the microring resonator.
It appears that this indeed happens in a vast region of the parameter space, mostly for bitrates up to \qty{500}{\Mbps}, and at every frequency detuning, although at different input powers.
Similarly, also the ``AND 2 with 1 R-bit'' task is solved by the RC network, but not at the same performance level as the previous task and for a limited set of parameters.
We can therefore summarize the results of the linear tasks by saying that the system is able to provide memory of 1 bit in the past in almost every configurations, memory of 2 bits in past for a restricted set of parameters, and is not able to provide memory of three or more bits in the past.
These considerations are validated by the analysis of the maps of $RB$, which verify that this memory comes indeed by the microring resonator response and not from nonlinearities already imprinted on the input optical signal by the EO conversion. The extension of the memory to just two bits in the past must be seen as the maximum intrinsic performance of the microresonator/bus waveguide system. It does not mean a binding limit. In fact, more memory can be provided a priori by modifying the encoding of virtual nodes \cite{Memory_Time}, or by exploiting a hybridized space-time approach  \cite{sunada2021photonic}. In this latter, one can vary the reservoir topology, e.g., by coupling spirals to the system, and thus, providing temporal memory between $n$ bits in the past. 

The nonlinear tasks, instead, require both memory and nonlinearity. Comparing the results of ``XOR 1 with 1 R-bit'' and ``XOR 1 with 2 R-bit'', we observe that the system struggles to deliver both.
In fact, if providing only the current bit to the ridge regression leads to a $\bber$ of around \ber{-1.7} and only for specific combination of the system parameters, with both bits the ridge regression is able to reach the statistical error limit on a wide region. Also in this case, best results are achieved when free carrier nonlinearities are used (i.e. negative detuning and bit rates of \qty{100}{\Mbps}).
Moreover, the same can be observed by comparing the maps of $RB$ for ``XOR 2'' with \numlist{1;2;3} R-bits. It is worth noticing that the experimental measurements are based on the nonlinear response encoded within a single signal of a pump laser. As a result, bits that have a zero value in the PRBS trace do not carry optical energy into the microresonator. Therefore, most of the virtual nodes associated with them are sampled within the background noise. This problem can be solved by exploiting a pump and probe experiment, and then, nonlinearly imprinting the information on a second signal \cite{Borghi_2021}.
However, in this work we are not interested in the absolute performance, but rather in the relative trend of the bit rate for the logical operations as a function of the number of bits supplied to the ridge regression in the training procedure.

These results show that both thermal and free carrier nonlinearities impact the performance of the microring reservoir.
Specifically, we can see a generalised improvement of RB for bitrates from \qtyrange{100}{400}{\Mbps}.
These values are close to the inverse of the free carrier lifetime $\sim 1/\tau_{FC} \sim \qtyrange{200}{500}{\MHz}$.
A similar scenario has already been reported by \cite{Borghi_2021}, where the higher free carrier lifetime of \qty{45}{\ns} gave rise to best performance around \qty{20}{\Mbps} in the XOR 1 with 1 R-bit task.
On the other hand, the interplay of free carrier and thermal nonlinearities, whose lifetime is $\sim 1/\tau_{th} \sim \qtyrange{10}{20}{\MHz}$,
do not deteriorate the device performance if the system is not in a bistable regime.
We cannot, however, observe the effect of the thermal nonlinearities alone as it would arise below \qty{20}{\Mbps}. It is good to point out that the photon lifetime, and therefore, the charge and discharge of the microresonator, does not provide memory between the bits.
In fact, its effect should occur around a frequency of $\sim \qty{200}{\GHz}$.

\section{Conclusion}
\label{sec:conclusion}

In this paper, a microresonator coupled to a bus waveguide is exploited as a reservoir in a RC network. The input information is encoded exploiting the temporal approach via virtual nodes, and thus, modulating the amplitude of a single pump laser. The system is trained offline through a ridge regression. By injecting into the bus waveguide a Pseudo Random Binary Sequence dependent on the bitrate, frequency, and average power, we studied linear and nonlinear logic operations, such as AND and XOR. The solution of the first tasks requires memory between the bits on which the logical operation occurs, while that of the second ones requires both: memory and nonlinearity understood as an activation function. Here, we show that a structure as simple as a single microring resonator can already exhibits a complex response, which can be harnessed to yield memory or to nonlinear transform the signal. In order to distinguish between the two features, we isolated the microring contributions to each of the two by providing external memory to the ridge regression and by testing linear tasks. 
The ridge regression succeeds in exploiting the nonlinearity of the microring resonator, either as an activation function, supplying the nonlinear transformation, or for memory, storing information of the past bits into the current one. 
Furthermore, the single microresonator can induce both memory and nonlinear activation function, overcoming the performance obtained by using the inter-symbolic interference of the input optical signal for tasks which consider the present bit and the past one.

Finally, we observed that, especially in photonic approaches relying on super-sampling and offline analysis, it is critical to know if the input optical signal already contains distortions, created during the modulation, which are enough for the ridge regression to solve the task.
Indeed, the optical modulator itself can distort the signal by imprinting spurious nonlinearities and inter-symbolic interference on the optical signal which can then be used by the ridge regression to get the target. Consequently, the virtual node approach requires the analysis of both experimental signals of the reservoir: the input and the output optical signals. Only the knowledge of both allows determining if the system under test provides the necessary nonlinearity to obtain the target. Therefore, it is important to analyze the $RB$ values. In other words, solving the task with just the experimental response of the output does not give a concrete proof of the reservoir performance. It only allows stating that the experimental apparatus as a whole system solves the task.

\section*{Acknowledgements}

The authors would like to thank Massimo Borghi for continuous and insightful exchanges and Enrico Moser for technical support.

\section*{Funding}
This project has received funding from the European Research Council (ERC) under the
European Union’s Horizon 2020 research and innovation programme (grant agreement No
788793, BACKUP), and from the MIUR under the project PRIN PELM (20177 PSCKT).

\section*{Disclosures}
The authors declare no conflicts of interest.

\bibliography{Bibliography.bib}

\begin{thebibliography}{33}%
\makeatletter
\providecommand \@ifxundefined [1]{%
 \@ifx{#1\undefined}
}%
\providecommand \@ifnum [1]{%
 \ifnum #1\expandafter \@firstoftwo
 \else \expandafter \@secondoftwo
 \fi
}%
\providecommand \@ifx [1]{%
 \ifx #1\expandafter \@firstoftwo
 \else \expandafter \@secondoftwo
 \fi
}%
\providecommand \natexlab [1]{#1}%
\providecommand \enquote  [1]{``#1''}%
\providecommand \bibnamefont  [1]{#1}%
\providecommand \bibfnamefont [1]{#1}%
\providecommand \citenamefont [1]{#1}%
\providecommand \href@noop [0]{\@secondoftwo}%
\providecommand \href [0]{\begingroup \@sanitize@url \@href}%
\providecommand \@href[1]{\@@startlink{#1}\@@href}%
\providecommand \@@href[1]{\endgroup#1\@@endlink}%
\providecommand \@sanitize@url [0]{\catcode `\\12\catcode `\$12\catcode
  `\&12\catcode `\#12\catcode `\^12\catcode `\_12\catcode `\%12\relax}%
\providecommand \@@startlink[1]{}%
\providecommand \@@endlink[0]{}%
\providecommand \url  [0]{\begingroup\@sanitize@url \@url }%
\providecommand \@url [1]{\endgroup\@href {#1}{\urlprefix }}%
\providecommand \urlprefix  [0]{URL }%
\providecommand \Eprint [0]{\href }%
\providecommand \doibase [0]{https://doi.org/}%
\providecommand \selectlanguage [0]{\@gobble}%
\providecommand \bibinfo  [0]{\@secondoftwo}%
\providecommand \bibfield  [0]{\@secondoftwo}%
\providecommand \translation [1]{[#1]}%
\providecommand \BibitemOpen [0]{}%
\providecommand \bibitemStop [0]{}%
\providecommand \bibitemNoStop [0]{.\EOS\space}%
\providecommand \EOS [0]{\spacefactor3000\relax}%
\providecommand \BibitemShut  [1]{\csname bibitem#1\endcsname}%
\let\auto@bib@innerbib\@empty
\bibitem [{\citenamefont {Thompson}\ \emph {et~al.}(2020)\citenamefont
  {Thompson}, \citenamefont {Greenewald}, \citenamefont {Lee},\ and\
  \citenamefont {Manso}}]{thompson2020computational}%
  \BibitemOpen
  \bibfield  {author} {\bibinfo {author} {\bibfnamefont {N.~C.}\ \bibnamefont
  {Thompson}}, \bibinfo {author} {\bibfnamefont {K.}~\bibnamefont
  {Greenewald}}, \bibinfo {author} {\bibfnamefont {K.}~\bibnamefont {Lee}},\
  and\ \bibinfo {author} {\bibfnamefont {G.~F.}\ \bibnamefont {Manso}},\
  }\bibfield  {title} {\bibinfo {title} {The computational limits of deep
  learning},\ }\href@noop {} {\bibfield  {journal} {\bibinfo  {journal} {arXiv
  preprint arXiv:2007.05558}\ } (\bibinfo {year} {2020})}\BibitemShut {NoStop}%
\bibitem [{\citenamefont {Chen}\ and\ \citenamefont {Lin}(2014)}]{chen2014big}%
  \BibitemOpen
  \bibfield  {author} {\bibinfo {author} {\bibfnamefont {X.-W.}\ \bibnamefont
  {Chen}}\ and\ \bibinfo {author} {\bibfnamefont {X.}~\bibnamefont {Lin}},\
  }\bibfield  {title} {\bibinfo {title} {Big data deep learning: challenges and
  perspectives},\ }\href@noop {} {\bibfield  {journal} {\bibinfo  {journal}
  {IEEE access}\ }\textbf {\bibinfo {volume} {2}},\ \bibinfo {pages} {514}
  (\bibinfo {year} {2014})}\BibitemShut {NoStop}%
\bibitem [{\citenamefont {Furber}(2016)}]{furber2016large}%
  \BibitemOpen
  \bibfield  {author} {\bibinfo {author} {\bibfnamefont {S.}~\bibnamefont
  {Furber}},\ }\bibfield  {title} {\bibinfo {title} {Large-scale neuromorphic
  computing systems},\ }\href@noop {} {\bibfield  {journal} {\bibinfo
  {journal} {Journal of neural engineering}\ }\textbf {\bibinfo {volume}
  {13}},\ \bibinfo {pages} {051001} (\bibinfo {year} {2016})}\BibitemShut
  {NoStop}%
\bibitem [{\citenamefont {Nakane}\ \emph {et~al.}(2018)\citenamefont {Nakane},
  \citenamefont {Tanaka},\ and\ \citenamefont {Hirose}}]{nakane2018reservoir}%
  \BibitemOpen
  \bibfield  {author} {\bibinfo {author} {\bibfnamefont {R.}~\bibnamefont
  {Nakane}}, \bibinfo {author} {\bibfnamefont {G.}~\bibnamefont {Tanaka}},\
  and\ \bibinfo {author} {\bibfnamefont {A.}~\bibnamefont {Hirose}},\
  }\bibfield  {title} {\bibinfo {title} {Reservoir computing with spin waves
  excited in a garnet film},\ }\href@noop {} {\bibfield  {journal} {\bibinfo
  {journal} {Ieee Access}\ }\textbf {\bibinfo {volume} {6}},\ \bibinfo {pages}
  {4462} (\bibinfo {year} {2018})}\BibitemShut {NoStop}%
\bibitem [{\citenamefont {Sunada}\ and\ \citenamefont
  {Uchida}(2021)}]{sunada2021photonic}%
  \BibitemOpen
  \bibfield  {author} {\bibinfo {author} {\bibfnamefont {S.}~\bibnamefont
  {Sunada}}\ and\ \bibinfo {author} {\bibfnamefont {A.}~\bibnamefont
  {Uchida}},\ }\bibfield  {title} {\bibinfo {title} {Photonic neural field on a
  silicon chip: large-scale, high-speed neuro-inspired computing and sensing},\
  }\href {https://doi.org/10.1364/OPTICA.434918} {\bibfield  {journal}
  {\bibinfo  {journal} {Optica}\ }\textbf {\bibinfo {volume} {8}},\ \bibinfo
  {pages} {1388} (\bibinfo {year} {2021})}\BibitemShut {NoStop}%
\bibitem [{\citenamefont {Du}\ \emph {et~al.}(2017)\citenamefont {Du},
  \citenamefont {Cai}, \citenamefont {Zidan}, \citenamefont {Ma}, \citenamefont
  {Lee},\ and\ \citenamefont {Lu}}]{du2017reservoir}%
  \BibitemOpen
  \bibfield  {author} {\bibinfo {author} {\bibfnamefont {C.}~\bibnamefont
  {Du}}, \bibinfo {author} {\bibfnamefont {F.}~\bibnamefont {Cai}}, \bibinfo
  {author} {\bibfnamefont {M.~A.}\ \bibnamefont {Zidan}}, \bibinfo {author}
  {\bibfnamefont {W.}~\bibnamefont {Ma}}, \bibinfo {author} {\bibfnamefont
  {S.~H.}\ \bibnamefont {Lee}},\ and\ \bibinfo {author} {\bibfnamefont {W.~D.}\
  \bibnamefont {Lu}},\ }\bibfield  {title} {\bibinfo {title} {Reservoir
  computing using dynamic memristors for temporal information processing},\
  }\href@noop {} {\bibfield  {journal} {\bibinfo  {journal} {Nature
  communications}\ }\textbf {\bibinfo {volume} {8}},\ \bibinfo {pages} {1}
  (\bibinfo {year} {2017})}\BibitemShut {NoStop}%
\bibitem [{\citenamefont {Merolla}\ \emph {et~al.}(2014)\citenamefont
  {Merolla}, \citenamefont {Arthur}, \citenamefont {Alvarez-Icaza},
  \citenamefont {Cassidy}, \citenamefont {Sawada}, \citenamefont {Akopyan},
  \citenamefont {Jackson}, \citenamefont {Imam}, \citenamefont {Guo},
  \citenamefont {Nakamura} \emph {et~al.}}]{merolla2014million}%
  \BibitemOpen
  \bibfield  {author} {\bibinfo {author} {\bibfnamefont {P.~A.}\ \bibnamefont
  {Merolla}}, \bibinfo {author} {\bibfnamefont {J.~V.}\ \bibnamefont {Arthur}},
  \bibinfo {author} {\bibfnamefont {R.}~\bibnamefont {Alvarez-Icaza}}, \bibinfo
  {author} {\bibfnamefont {A.~S.}\ \bibnamefont {Cassidy}}, \bibinfo {author}
  {\bibfnamefont {J.}~\bibnamefont {Sawada}}, \bibinfo {author} {\bibfnamefont
  {F.}~\bibnamefont {Akopyan}}, \bibinfo {author} {\bibfnamefont {B.~L.}\
  \bibnamefont {Jackson}}, \bibinfo {author} {\bibfnamefont {N.}~\bibnamefont
  {Imam}}, \bibinfo {author} {\bibfnamefont {C.}~\bibnamefont {Guo}}, \bibinfo
  {author} {\bibfnamefont {Y.}~\bibnamefont {Nakamura}}, \emph {et~al.},\
  }\bibfield  {title} {\bibinfo {title} {A million spiking-neuron integrated
  circuit with a scalable communication network and interface},\ }\href@noop {}
  {\bibfield  {journal} {\bibinfo  {journal} {Science}\ }\textbf {\bibinfo
  {volume} {345}},\ \bibinfo {pages} {668} (\bibinfo {year}
  {2014})}\BibitemShut {NoStop}%
\bibitem [{\citenamefont {Guo}\ \emph {et~al.}(2019)\citenamefont {Guo},
  \citenamefont {Xiang}, \citenamefont {Zhang}, \citenamefont {Lin},
  \citenamefont {Wen},\ and\ \citenamefont {Hao}}]{guo2019polarization}%
  \BibitemOpen
  \bibfield  {author} {\bibinfo {author} {\bibfnamefont {X.~X.}\ \bibnamefont
  {Guo}}, \bibinfo {author} {\bibfnamefont {S.~Y.}\ \bibnamefont {Xiang}},
  \bibinfo {author} {\bibfnamefont {Y.~H.}\ \bibnamefont {Zhang}}, \bibinfo
  {author} {\bibfnamefont {L.}~\bibnamefont {Lin}}, \bibinfo {author}
  {\bibfnamefont {A.~J.}\ \bibnamefont {Wen}},\ and\ \bibinfo {author}
  {\bibfnamefont {Y.}~\bibnamefont {Hao}},\ }\bibfield  {title} {\bibinfo
  {title} {Polarization multiplexing reservoir computing based on a vcsel with
  polarized optical feedback},\ }\href@noop {} {\bibfield  {journal} {\bibinfo
  {journal} {IEEE Journal of Selected Topics in Quantum Electronics}\ }\textbf
  {\bibinfo {volume} {26}},\ \bibinfo {pages} {1} (\bibinfo {year}
  {2019})}\BibitemShut {NoStop}%
\bibitem [{\citenamefont {Wu}\ \emph {et~al.}(2021)\citenamefont {Wu},
  \citenamefont {Yu}, \citenamefont {Lee}, \citenamefont {Peng}, \citenamefont
  {Takeuchi},\ and\ \citenamefont {Li}}]{wu2021programmable}%
  \BibitemOpen
  \bibfield  {author} {\bibinfo {author} {\bibfnamefont {C.}~\bibnamefont
  {Wu}}, \bibinfo {author} {\bibfnamefont {H.}~\bibnamefont {Yu}}, \bibinfo
  {author} {\bibfnamefont {S.}~\bibnamefont {Lee}}, \bibinfo {author}
  {\bibfnamefont {R.}~\bibnamefont {Peng}}, \bibinfo {author} {\bibfnamefont
  {I.}~\bibnamefont {Takeuchi}},\ and\ \bibinfo {author} {\bibfnamefont
  {M.}~\bibnamefont {Li}},\ }\bibfield  {title} {\bibinfo {title} {Programmable
  phase-change metasurfaces on waveguides for multimode photonic convolutional
  neural network},\ }\href@noop {} {\bibfield  {journal} {\bibinfo  {journal}
  {Nature communications}\ }\textbf {\bibinfo {volume} {12}},\ \bibinfo {pages}
  {1} (\bibinfo {year} {2021})}\BibitemShut {NoStop}%
\bibitem [{\citenamefont {Jaeger}(2001)}]{jaeger2001echo}%
  \BibitemOpen
  \bibfield  {author} {\bibinfo {author} {\bibfnamefont {H.}~\bibnamefont
  {Jaeger}},\ }\bibfield  {title} {\bibinfo {title} {The “echo state”
  approach to analysing and training recurrent neural networks-with an erratum
  note},\ }\href@noop {} {\bibfield  {journal} {\bibinfo  {journal} {{Bonn,
  Germany: German National Research Center for Information Technology GMD
  Technical Report}}\ }\textbf {\bibinfo {volume} {148}},\ \bibinfo {pages}
  {13} (\bibinfo {year} {2001})}\BibitemShut {NoStop}%
\bibitem [{\citenamefont {Maass}\ and\ \citenamefont
  {Markram}(2004)}]{maass2004computational}%
  \BibitemOpen
  \bibfield  {author} {\bibinfo {author} {\bibfnamefont {W.}~\bibnamefont
  {Maass}}\ and\ \bibinfo {author} {\bibfnamefont {H.}~\bibnamefont
  {Markram}},\ }\bibfield  {title} {\bibinfo {title} {On the computational
  power of circuits of spiking neurons},\ }\href@noop {} {\bibfield  {journal}
  {\bibinfo  {journal} {J. Comput. Syst. Sci.}\ }\textbf {\bibinfo {volume}
  {69}},\ \bibinfo {pages} {593} (\bibinfo {year} {2004})}\BibitemShut
  {NoStop}%
\bibitem [{\citenamefont {De~Marinis}\ \emph {et~al.}(2019)\citenamefont
  {De~Marinis}, \citenamefont {Cococcioni}, \citenamefont {Castoldi},\ and\
  \citenamefont {Andriolli}}]{Survey}%
  \BibitemOpen
  \bibfield  {author} {\bibinfo {author} {\bibfnamefont {L.}~\bibnamefont
  {De~Marinis}}, \bibinfo {author} {\bibfnamefont {M.}~\bibnamefont
  {Cococcioni}}, \bibinfo {author} {\bibfnamefont {P.}~\bibnamefont
  {Castoldi}},\ and\ \bibinfo {author} {\bibfnamefont {N.}~\bibnamefont
  {Andriolli}},\ }\bibfield  {title} {\bibinfo {title} {Photonic neural
  networks: A survey},\ }\href {https://doi.org/10.1109/ACCESS.2019.2957245}
  {\bibfield  {journal} {\bibinfo  {journal} {IEEE Access}\ }\textbf {\bibinfo
  {volume} {7}},\ \bibinfo {pages} {175827} (\bibinfo {year}
  {2019})}\BibitemShut {NoStop}%
\bibitem [{\citenamefont {Coarer}\ \emph {et~al.}(2018)\citenamefont {Coarer},
  \citenamefont {Sciamanna}, \citenamefont {Katumba}, \citenamefont
  {Freiberger}, \citenamefont {Dambre}, \citenamefont {Bienstman},\ and\
  \citenamefont {Rontani}}]{IEEE_Bienstman}%
  \BibitemOpen
  \bibfield  {author} {\bibinfo {author} {\bibfnamefont {F.~D.}\ \bibnamefont
  {Coarer}}, \bibinfo {author} {\bibfnamefont {M.}~\bibnamefont {Sciamanna}},
  \bibinfo {author} {\bibfnamefont {A.}~\bibnamefont {Katumba}}, \bibinfo
  {author} {\bibfnamefont {M.}~\bibnamefont {Freiberger}}, \bibinfo {author}
  {\bibfnamefont {J.}~\bibnamefont {Dambre}}, \bibinfo {author} {\bibfnamefont
  {P.}~\bibnamefont {Bienstman}},\ and\ \bibinfo {author} {\bibfnamefont
  {D.}~\bibnamefont {Rontani}},\ }\bibfield  {title} {\bibinfo {title}
  {All-optical reservoir computing on a photonic chip using silicon-based ring
  resonators},\ }\href {https://doi.org/10.1109/JSTQE.2018.2836985} {\bibfield
  {journal} {\bibinfo  {journal} {IEEE Journal of Selected Topics in Quantum
  Electronics}\ }\textbf {\bibinfo {volume} {24}},\ \bibinfo {pages} {1}
  (\bibinfo {year} {2018})}\BibitemShut {NoStop}%
\bibitem [{\citenamefont {Appeltant}\ \emph {et~al.}(2011)\citenamefont
  {Appeltant}, \citenamefont {Soriano}, \citenamefont {Van~der Sande},
  \citenamefont {Danckaert}, \citenamefont {Massar}, \citenamefont {Dambre},
  \citenamefont {Schrauwen}, \citenamefont {Mirasso},\ and\ \citenamefont
  {Fischer}}]{appeltant2011information}%
  \BibitemOpen
  \bibfield  {author} {\bibinfo {author} {\bibfnamefont {L.}~\bibnamefont
  {Appeltant}}, \bibinfo {author} {\bibfnamefont {M.~C.}\ \bibnamefont
  {Soriano}}, \bibinfo {author} {\bibfnamefont {G.}~\bibnamefont {Van~der
  Sande}}, \bibinfo {author} {\bibfnamefont {J.}~\bibnamefont {Danckaert}},
  \bibinfo {author} {\bibfnamefont {S.}~\bibnamefont {Massar}}, \bibinfo
  {author} {\bibfnamefont {J.}~\bibnamefont {Dambre}}, \bibinfo {author}
  {\bibfnamefont {B.}~\bibnamefont {Schrauwen}}, \bibinfo {author}
  {\bibfnamefont {C.~R.}\ \bibnamefont {Mirasso}},\ and\ \bibinfo {author}
  {\bibfnamefont {I.}~\bibnamefont {Fischer}},\ }\bibfield  {title} {\bibinfo
  {title} {Information processing using a single dynamical node as complex
  system},\ }\href@noop {} {\bibfield  {journal} {\bibinfo  {journal} {Nature
  communications}\ }\textbf {\bibinfo {volume} {2}},\ \bibinfo {pages} {1}
  (\bibinfo {year} {2011})}\BibitemShut {NoStop}%
\bibitem [{\citenamefont {Donati}\ \emph {et~al.}(2021)\citenamefont {Donati},
  \citenamefont {Mirasso}, \citenamefont {Mancinelli}, \citenamefont {Pavesi},\
  and\ \citenamefont {Argyris}}]{donati2021microring}%
  \BibitemOpen
  \bibfield  {author} {\bibinfo {author} {\bibfnamefont {G.}~\bibnamefont
  {Donati}}, \bibinfo {author} {\bibfnamefont {C.~R.}\ \bibnamefont {Mirasso}},
  \bibinfo {author} {\bibfnamefont {M.}~\bibnamefont {Mancinelli}}, \bibinfo
  {author} {\bibfnamefont {L.}~\bibnamefont {Pavesi}},\ and\ \bibinfo {author}
  {\bibfnamefont {A.}~\bibnamefont {Argyris}},\ }\href@noop {} {\bibinfo
  {title} {Microring resonators with external optical feedback for time delay
  reservoir computing}} (\bibinfo {year} {2021}),\ \Eprint
  {https://arxiv.org/abs/2109.11486} {arXiv:2109.11486 [physics.optics]}
  \BibitemShut {NoStop}%
\bibitem [{\citenamefont {Borghi}\ \emph
  {et~al.}(2021{\natexlab{a}})\citenamefont {Borghi}, \citenamefont {Biasi},\
  and\ \citenamefont {Pavesi}}]{Borghi_2021}%
  \BibitemOpen
  \bibfield  {author} {\bibinfo {author} {\bibfnamefont {M.}~\bibnamefont
  {Borghi}}, \bibinfo {author} {\bibfnamefont {S.}~\bibnamefont {Biasi}},\ and\
  \bibinfo {author} {\bibfnamefont {L.}~\bibnamefont {Pavesi}},\ }\bibfield
  {title} {\bibinfo {title} {Reservoir computing based on a silicon microring
  and time multiplexing for binary and analog operations},\ }\bibfield
  {journal} {\bibinfo  {journal} {Scientific Reports}\ }\textbf {\bibinfo
  {volume} {11}},\ \href {https://doi.org/10.1038/s41598-021-94952-5}
  {10.1038/s41598-021-94952-5} (\bibinfo {year}
  {2021}{\natexlab{a}})\BibitemShut {NoStop}%
\bibitem [{\citenamefont {Borghi}\ \emph
  {et~al.}(2021{\natexlab{b}})\citenamefont {Borghi}, \citenamefont
  {Bazzanella}, \citenamefont {Mancinelli},\ and\ \citenamefont
  {Pavesi}}]{borghi2020OntheModeling}%
  \BibitemOpen
  \bibfield  {author} {\bibinfo {author} {\bibfnamefont {M.}~\bibnamefont
  {Borghi}}, \bibinfo {author} {\bibfnamefont {D.}~\bibnamefont {Bazzanella}},
  \bibinfo {author} {\bibfnamefont {M.}~\bibnamefont {Mancinelli}},\ and\
  \bibinfo {author} {\bibfnamefont {L.}~\bibnamefont {Pavesi}},\ }\bibfield
  {title} {\bibinfo {title} {On the modeling of thermal and free carrier
  nonlinearities in silicon-on-insulator microring resonators},\ }\href@noop {}
  {\bibfield  {journal} {\bibinfo  {journal} {Optics Express}\ }\textbf
  {\bibinfo {volume} {29}},\ \bibinfo {pages} {4363} (\bibinfo {year}
  {2021}{\natexlab{b}})}\BibitemShut {NoStop}%
\bibitem [{\citenamefont {Harkhoe}\ \emph {et~al.}(2020)\citenamefont
  {Harkhoe}, \citenamefont {Verschaffelt}, \citenamefont {Katumba},
  \citenamefont {Bienstman},\ and\ \citenamefont {der Sande}}]{Memory_OSA}%
  \BibitemOpen
  \bibfield  {author} {\bibinfo {author} {\bibfnamefont {K.}~\bibnamefont
  {Harkhoe}}, \bibinfo {author} {\bibfnamefont {G.}~\bibnamefont
  {Verschaffelt}}, \bibinfo {author} {\bibfnamefont {A.}~\bibnamefont
  {Katumba}}, \bibinfo {author} {\bibfnamefont {P.}~\bibnamefont {Bienstman}},\
  and\ \bibinfo {author} {\bibfnamefont {G.~V.}\ \bibnamefont {der Sande}},\
  }\bibfield  {title} {\bibinfo {title} {Demonstrating delay-based reservoir
  computing using a compact photonic integrated chip},\ }\href
  {https://doi.org/10.1364/OE.382556} {\bibfield  {journal} {\bibinfo
  {journal} {Opt. Express}\ }\textbf {\bibinfo {volume} {28}},\ \bibinfo
  {pages} {3086} (\bibinfo {year} {2020})}\BibitemShut {NoStop}%
\bibitem [{\citenamefont {Bueno}\ \emph {et~al.}(2017)\citenamefont {Bueno},
  \citenamefont {Brunner}, \citenamefont {Soriano},\ and\ \citenamefont
  {Fischer}}]{Time_Fischer}%
  \BibitemOpen
  \bibfield  {author} {\bibinfo {author} {\bibfnamefont {J.}~\bibnamefont
  {Bueno}}, \bibinfo {author} {\bibfnamefont {D.}~\bibnamefont {Brunner}},
  \bibinfo {author} {\bibfnamefont {M.~C.}\ \bibnamefont {Soriano}},\ and\
  \bibinfo {author} {\bibfnamefont {I.}~\bibnamefont {Fischer}},\ }\bibfield
  {title} {\bibinfo {title} {Conditions for reservoir computing performance
  using semiconductor lasers with delayed optical feedback},\ }\href
  {https://doi.org/10.1364/OE.25.002401} {\bibfield  {journal} {\bibinfo
  {journal} {Opt. Express}\ }\textbf {\bibinfo {volume} {25}},\ \bibinfo
  {pages} {2401} (\bibinfo {year} {2017})}\BibitemShut {NoStop}%
\bibitem [{\citenamefont {Vandoorne}\ \emph {et~al.}(2014)\citenamefont
  {Vandoorne}, \citenamefont {Mechet}, \citenamefont {Van~Vaerenbergh},
  \citenamefont {Fiers}, \citenamefont {Morthier}, \citenamefont {Verstraeten},
  \citenamefont {Schrauwen}, \citenamefont {Dambre},\ and\ \citenamefont
  {Bienstman}}]{NatCom2014}%
  \BibitemOpen
  \bibfield  {author} {\bibinfo {author} {\bibfnamefont {K.}~\bibnamefont
  {Vandoorne}}, \bibinfo {author} {\bibfnamefont {P.}~\bibnamefont {Mechet}},
  \bibinfo {author} {\bibfnamefont {T.}~\bibnamefont {Van~Vaerenbergh}},
  \bibinfo {author} {\bibfnamefont {M.}~\bibnamefont {Fiers}}, \bibinfo
  {author} {\bibfnamefont {G.}~\bibnamefont {Morthier}}, \bibinfo {author}
  {\bibfnamefont {D.}~\bibnamefont {Verstraeten}}, \bibinfo {author}
  {\bibfnamefont {B.}~\bibnamefont {Schrauwen}}, \bibinfo {author}
  {\bibfnamefont {J.}~\bibnamefont {Dambre}},\ and\ \bibinfo {author}
  {\bibfnamefont {P.}~\bibnamefont {Bienstman}},\ }\bibfield  {title} {\bibinfo
  {title} {Experimental demonstration of reservoir computing on a silicon
  photonics chip},\ }\href {https://doi.org/10.1038/ncomms4541} {\bibfield
  {journal} {\bibinfo  {journal} {Nature Communications}\ }\textbf {\bibinfo
  {volume} {5}},\ \bibinfo {pages} {3541} (\bibinfo {year} {2014})}\BibitemShut
  {NoStop}%
\bibitem [{\citenamefont {Johnson}\ \emph {et~al.}(2006)\citenamefont
  {Johnson}, \citenamefont {Borselli},\ and\ \citenamefont
  {Painter}}]{johnson2006self}%
  \BibitemOpen
  \bibfield  {author} {\bibinfo {author} {\bibfnamefont {T.~J.}\ \bibnamefont
  {Johnson}}, \bibinfo {author} {\bibfnamefont {M.}~\bibnamefont {Borselli}},\
  and\ \bibinfo {author} {\bibfnamefont {O.}~\bibnamefont {Painter}},\
  }\bibfield  {title} {\bibinfo {title} {Self-induced optical modulation of the
  transmission through a high-q silicon microdisk resonator},\ }\href@noop {}
  {\bibfield  {journal} {\bibinfo  {journal} {Optics Express}\ }\textbf
  {\bibinfo {volume} {14}},\ \bibinfo {pages} {817} (\bibinfo {year}
  {2006})}\BibitemShut {NoStop}%
\bibitem [{\citenamefont {Baker}\ \emph {et~al.}(2012)\citenamefont {Baker},
  \citenamefont {Stapfner}, \citenamefont {Parrain}, \citenamefont {Ducci},
  \citenamefont {Leo}, \citenamefont {Weig},\ and\ \citenamefont
  {Favero}}]{self_Baker}%
  \BibitemOpen
  \bibfield  {author} {\bibinfo {author} {\bibfnamefont {C.}~\bibnamefont
  {Baker}}, \bibinfo {author} {\bibfnamefont {S.}~\bibnamefont {Stapfner}},
  \bibinfo {author} {\bibfnamefont {D.}~\bibnamefont {Parrain}}, \bibinfo
  {author} {\bibfnamefont {S.}~\bibnamefont {Ducci}}, \bibinfo {author}
  {\bibfnamefont {G.}~\bibnamefont {Leo}}, \bibinfo {author} {\bibfnamefont
  {E.~M.}\ \bibnamefont {Weig}},\ and\ \bibinfo {author} {\bibfnamefont
  {I.}~\bibnamefont {Favero}},\ }\bibfield  {title} {\bibinfo {title} {Optical
  instability and self-pulsing in silicon nitride whispering gallery
  resonators},\ }\href {https://doi.org/10.1364/OE.20.029076} {\bibfield
  {journal} {\bibinfo  {journal} {Opt. Express}\ }\textbf {\bibinfo {volume}
  {20}},\ \bibinfo {pages} {29076} (\bibinfo {year} {2012})}\BibitemShut
  {NoStop}%
\bibitem [{\citenamefont {Tian}\ \emph {et~al.}(2013)\citenamefont {Tian},
  \citenamefont {Zhang},\ and\ \citenamefont {Yang}}]{XOR_coupled-resonator}%
  \BibitemOpen
  \bibfield  {author} {\bibinfo {author} {\bibfnamefont {Y.}~\bibnamefont
  {Tian}}, \bibinfo {author} {\bibfnamefont {L.}~\bibnamefont {Zhang}},\ and\
  \bibinfo {author} {\bibfnamefont {L.}~\bibnamefont {Yang}},\ }\bibfield
  {title} {\bibinfo {title} {{XOR/XNOR directed logic circuit based on
  coupled-resonator-induced transparency}},\ }in\ \href
  {https://doi.org/10.1117/12.2028584} {\emph {\bibinfo {booktitle} {Optics and
  Photonics for Information Processing VII}}},\ Vol.\ \bibinfo {volume}
  {8855},\ \bibinfo {editor} {edited by\ \bibinfo {editor} {\bibfnamefont
  {K.~M.}\ \bibnamefont {Iftekharuddin}}, \bibinfo {editor} {\bibfnamefont
  {A.~A.~S.}\ \bibnamefont {Awwal}},\ and\ \bibinfo {editor} {\bibfnamefont
  {A.}~\bibnamefont {Márquez}}},\ \bibinfo {organization} {International
  Society for Optics and Photonics}\ (\bibinfo  {publisher} {SPIE},\ \bibinfo
  {year} {2013})\ pp.\ \bibinfo {pages} {187 -- 192}\BibitemShut {NoStop}%
\bibitem [{\citenamefont {Kim}\ \emph {et~al.}(2006)\citenamefont {Kim},
  \citenamefont {Kang}, \citenamefont {Kim},\ and\ \citenamefont
  {Han}}]{XOR_Han}%
  \BibitemOpen
  \bibfield  {author} {\bibinfo {author} {\bibfnamefont {J.-Y.}\ \bibnamefont
  {Kim}}, \bibinfo {author} {\bibfnamefont {J.-M.}\ \bibnamefont {Kang}},
  \bibinfo {author} {\bibfnamefont {T.-Y.}\ \bibnamefont {Kim}},\ and\ \bibinfo
  {author} {\bibfnamefont {S.-K.}\ \bibnamefont {Han}},\ }\bibfield  {title}
  {\bibinfo {title} {All-optical multiple logic gates with xor, nor, or, and
  nand functions using parallel soa-mzi structures: theory and experiment},\
  }\href {https://doi.org/10.1109/JLT.2006.880593} {\bibfield  {journal}
  {\bibinfo  {journal} {Journal of Lightwave Technology}\ }\textbf {\bibinfo
  {volume} {24}},\ \bibinfo {pages} {3392} (\bibinfo {year}
  {2006})}\BibitemShut {NoStop}%
\bibitem [{\citenamefont {Tian}\ \emph {et~al.}(2016)\citenamefont {Tian},
  \citenamefont {Li}, \citenamefont {Liu}, \citenamefont {Xiao}, \citenamefont
  {Zhao}, \citenamefont {Yang}, \citenamefont {Zhao}, \citenamefont {Han},\
  and\ \citenamefont {Gao}}]{Xor_tworingNew}%
  \BibitemOpen
  \bibfield  {author} {\bibinfo {author} {\bibfnamefont {Y.}~\bibnamefont
  {Tian}}, \bibinfo {author} {\bibfnamefont {D.}~\bibnamefont {Li}}, \bibinfo
  {author} {\bibfnamefont {Z.}~\bibnamefont {Liu}}, \bibinfo {author}
  {\bibfnamefont {H.}~\bibnamefont {Xiao}}, \bibinfo {author} {\bibfnamefont
  {G.}~\bibnamefont {Zhao}}, \bibinfo {author} {\bibfnamefont {J.}~\bibnamefont
  {Yang}}, \bibinfo {author} {\bibfnamefont {Y.}~\bibnamefont {Zhao}}, \bibinfo
  {author} {\bibfnamefont {G.}~\bibnamefont {Han}},\ and\ \bibinfo {author}
  {\bibfnamefont {X.}~\bibnamefont {Gao}},\ }\bibfield  {title} {\bibinfo
  {title} {Simulation and demonstration of directed xor/xnor logic gates using
  two cascaded microring resonators},\ }\href
  {https://doi.org/10.1109/JPHOT.2016.2535910} {\bibfield  {journal} {\bibinfo
  {journal} {IEEE Photonics Journal}\ }\textbf {\bibinfo {volume} {8}},\
  \bibinfo {pages} {1} (\bibinfo {year} {2016})}\BibitemShut {NoStop}%
\bibitem [{\citenamefont {Biasi}\ \emph {et~al.}(2019)\citenamefont {Biasi},
  \citenamefont {Guillemé}, \citenamefont {Volpini}, \citenamefont {Fontana},\
  and\ \citenamefont {Pavesi}}]{Time_Biasi}%
  \BibitemOpen
  \bibfield  {author} {\bibinfo {author} {\bibfnamefont {S.}~\bibnamefont
  {Biasi}}, \bibinfo {author} {\bibfnamefont {P.}~\bibnamefont {Guillemé}},
  \bibinfo {author} {\bibfnamefont {A.}~\bibnamefont {Volpini}}, \bibinfo
  {author} {\bibfnamefont {G.}~\bibnamefont {Fontana}},\ and\ \bibinfo {author}
  {\bibfnamefont {L.}~\bibnamefont {Pavesi}},\ }\bibfield  {title} {\bibinfo
  {title} {Time response of a microring resonator to a rectangular pulse in
  different coupling regimes},\ }\href
  {https://doi.org/10.1109/JLT.2019.2928640} {\bibfield  {journal} {\bibinfo
  {journal} {Journal of Lightwave Technology}\ }\textbf {\bibinfo {volume}
  {37}},\ \bibinfo {pages} {5091} (\bibinfo {year} {2019})}\BibitemShut
  {NoStop}%
\bibitem [{\citenamefont {Pernice}\ \emph {et~al.}(2011)\citenamefont
  {Pernice}, \citenamefont {Schuck}, \citenamefont {Li},\ and\ \citenamefont
  {Tang}}]{1.9ns_Pernice}%
  \BibitemOpen
  \bibfield  {author} {\bibinfo {author} {\bibfnamefont {W.~H.~P.}\
  \bibnamefont {Pernice}}, \bibinfo {author} {\bibfnamefont {C.}~\bibnamefont
  {Schuck}}, \bibinfo {author} {\bibfnamefont {M.}~\bibnamefont {Li}},\ and\
  \bibinfo {author} {\bibfnamefont {H.~X.}\ \bibnamefont {Tang}},\ }\bibfield
  {title} {\bibinfo {title} {Carrier and thermal dynamics of silicon photonic
  resonators at cryogenic temperatures},\ }\href
  {https://doi.org/10.1364/OE.19.003290} {\bibfield  {journal} {\bibinfo
  {journal} {Opt. Express}\ }\textbf {\bibinfo {volume} {19}},\ \bibinfo
  {pages} {3290} (\bibinfo {year} {2011})}\BibitemShut {NoStop}%
\bibitem [{\citenamefont {Xu}\ and\ \citenamefont {Lipson}(2007)}]{xu2007all}%
  \BibitemOpen
  \bibfield  {author} {\bibinfo {author} {\bibfnamefont {Q.}~\bibnamefont
  {Xu}}\ and\ \bibinfo {author} {\bibfnamefont {M.}~\bibnamefont {Lipson}},\
  }\bibfield  {title} {\bibinfo {title} {All-optical logic based on silicon
  micro-ring resonators},\ }\href@noop {} {\bibfield  {journal} {\bibinfo
  {journal} {Optics express}\ }\textbf {\bibinfo {volume} {15}},\ \bibinfo
  {pages} {924} (\bibinfo {year} {2007})}\BibitemShut {NoStop}%
\bibitem [{\citenamefont {Almeida}\ \emph {et~al.}(2004)\citenamefont
  {Almeida}, \citenamefont {Barrios}, \citenamefont {Panepucci},\ and\
  \citenamefont {Lipson}}]{almeida2004all}%
  \BibitemOpen
  \bibfield  {author} {\bibinfo {author} {\bibfnamefont {V.~R.}\ \bibnamefont
  {Almeida}}, \bibinfo {author} {\bibfnamefont {C.~A.}\ \bibnamefont
  {Barrios}}, \bibinfo {author} {\bibfnamefont {R.~R.}\ \bibnamefont
  {Panepucci}},\ and\ \bibinfo {author} {\bibfnamefont {M.}~\bibnamefont
  {Lipson}},\ }\bibfield  {title} {\bibinfo {title} {All-optical control of
  light on a silicon chip},\ }\href@noop {} {\bibfield  {journal} {\bibinfo
  {journal} {Nature}\ }\textbf {\bibinfo {volume} {431}},\ \bibinfo {pages}
  {1081} (\bibinfo {year} {2004})}\BibitemShut {NoStop}%
\bibitem [{\citenamefont {Luo}\ \emph {et~al.}(2012)\citenamefont {Luo},
  \citenamefont {Wiederhecker}, \citenamefont {Preston},\ and\ \citenamefont
  {Lipson}}]{luo2012power}%
  \BibitemOpen
  \bibfield  {author} {\bibinfo {author} {\bibfnamefont {L.-W.}\ \bibnamefont
  {Luo}}, \bibinfo {author} {\bibfnamefont {G.~S.}\ \bibnamefont
  {Wiederhecker}}, \bibinfo {author} {\bibfnamefont {K.}~\bibnamefont
  {Preston}},\ and\ \bibinfo {author} {\bibfnamefont {M.}~\bibnamefont
  {Lipson}},\ }\bibfield  {title} {\bibinfo {title} {Power insensitive silicon
  microring resonators},\ }\href@noop {} {\bibfield  {journal} {\bibinfo
  {journal} {Optics letters}\ }\textbf {\bibinfo {volume} {37}},\ \bibinfo
  {pages} {590} (\bibinfo {year} {2012})}\BibitemShut {NoStop}%
\bibitem [{\citenamefont {Van~Vaerenbergh}\ \emph {et~al.}(2012)\citenamefont
  {Van~Vaerenbergh}, \citenamefont {Fiers}, \citenamefont {Dambre},\ and\
  \citenamefont {Bienstman}}]{Thermal_Van}%
  \BibitemOpen
  \bibfield  {author} {\bibinfo {author} {\bibfnamefont {T.}~\bibnamefont
  {Van~Vaerenbergh}}, \bibinfo {author} {\bibfnamefont {M.}~\bibnamefont
  {Fiers}}, \bibinfo {author} {\bibfnamefont {J.}~\bibnamefont {Dambre}},\ and\
  \bibinfo {author} {\bibfnamefont {P.}~\bibnamefont {Bienstman}},\ }\bibfield
  {title} {\bibinfo {title} {Simplified description of self-pulsation and
  excitability by thermal and free-carrier effects in semiconductor
  microcavities},\ }\href {https://doi.org/10.1103/PhysRevA.86.063808}
  {\bibfield  {journal} {\bibinfo  {journal} {Phys. Rev. A}\ }\textbf {\bibinfo
  {volume} {86}},\ \bibinfo {pages} {063808} (\bibinfo {year}
  {2012})}\BibitemShut {NoStop}%
\bibitem [{\citenamefont {Vaerenbergh}\ \emph {et~al.}(2012)\citenamefont
  {Vaerenbergh}, \citenamefont {Fiers}, \citenamefont {Mechet}, \citenamefont
  {Spuesens}, \citenamefont {Kumar}, \citenamefont {Morthier}, \citenamefont
  {Schrauwen}, \citenamefont {Dambre},\ and\ \citenamefont
  {Bienstman}}]{Thermal_Van1}%
  \BibitemOpen
  \bibfield  {author} {\bibinfo {author} {\bibfnamefont {T.~V.}\ \bibnamefont
  {Vaerenbergh}}, \bibinfo {author} {\bibfnamefont {M.}~\bibnamefont {Fiers}},
  \bibinfo {author} {\bibfnamefont {P.}~\bibnamefont {Mechet}}, \bibinfo
  {author} {\bibfnamefont {T.}~\bibnamefont {Spuesens}}, \bibinfo {author}
  {\bibfnamefont {R.}~\bibnamefont {Kumar}}, \bibinfo {author} {\bibfnamefont
  {G.}~\bibnamefont {Morthier}}, \bibinfo {author} {\bibfnamefont
  {B.}~\bibnamefont {Schrauwen}}, \bibinfo {author} {\bibfnamefont
  {J.}~\bibnamefont {Dambre}},\ and\ \bibinfo {author} {\bibfnamefont
  {P.}~\bibnamefont {Bienstman}},\ }\bibfield  {title} {\bibinfo {title}
  {Cascadable excitability in microrings},\ }\href
  {https://doi.org/10.1364/OE.20.020292} {\bibfield  {journal} {\bibinfo
  {journal} {Opt. Express}\ }\textbf {\bibinfo {volume} {20}},\ \bibinfo
  {pages} {20292} (\bibinfo {year} {2012})}\BibitemShut {NoStop}%
\bibitem [{\citenamefont {Ort{\'i}­n}\ \emph {et~al.}(2015)\citenamefont
  {Ort{\'i}­n}, \citenamefont {Soriano}, \citenamefont {Pesquera},
  \citenamefont {Brunner}, \citenamefont {San-Mart{\'i}­n}, \citenamefont
  {Fischer}, \citenamefont {Mirasso},\ and\ \citenamefont
  {Guti{\'e}rrez}}]{Memory_Time}%
  \BibitemOpen
  \bibfield  {author} {\bibinfo {author} {\bibfnamefont {S.}~\bibnamefont
  {Ort{\'i}­n}}, \bibinfo {author} {\bibfnamefont {M.~C.}\ \bibnamefont
  {Soriano}}, \bibinfo {author} {\bibfnamefont {L.}~\bibnamefont {Pesquera}},
  \bibinfo {author} {\bibfnamefont {D.}~\bibnamefont {Brunner}}, \bibinfo
  {author} {\bibfnamefont {D.}~\bibnamefont {San-Mart{\'i}­n}}, \bibinfo
  {author} {\bibfnamefont {I.}~\bibnamefont {Fischer}}, \bibinfo {author}
  {\bibfnamefont {C.~R.}\ \bibnamefont {Mirasso}},\ and\ \bibinfo {author}
  {\bibfnamefont {J.~M.}\ \bibnamefont {Guti{\'e}rrez}},\ }\bibfield  {title}
  {\bibinfo {title} {A unified framework for reservoir computing and extreme
  learning machines based on a single time-delayed neuron},\ }\href
  {https://doi.org/10.1038/srep14945} {\bibfield  {journal} {\bibinfo
  {journal} {Scientific Rep.}\ }\textbf {\bibinfo {volume} {5}},\ \bibinfo
  {pages} {14945} (\bibinfo {year} {2015})}\BibitemShut {NoStop}%
\end{thebibliography}%

\end{document}